\documentclass[11pt, aip]{revtex4-1}

\usepackage{graphicx,rotating,subfigure}
\usepackage{amsmath}
\usepackage{pstricks}
\renewcommand{\vec}[1]{\mathbf{#1}}

\begin{document}

\title{Non-Newtonian behavior and molecular structure of Cooee bitumen under shear flow: a non-equilibrium molecular dynamics study}

\author{Claire A. Lemarchand}
\email{clairel@ruc.dk}
\affiliation{
  DNRF Centre ``Glass and Time'', IMFUFA,
  Department of Sciences, 
  Roskilde University, Postbox 260,
  DK-4000 Roskilde, Denmark $\mbox{}$
}

\author{Nicholas P. Bailey}
\affiliation{
  DNRF Centre ``Glass and Time'', IMFUFA,
  Department of Sciences,
  Roskilde University, Postbox 260,
  DK-4000 Roskilde, Denmark $\mbox{} $ $\mbox{} \mbox{}$
}

\author{Billy D. Todd}
\affiliation{
Department of Mathematics, 
Faculty of Science, Engineering and Technology, and Centre for Molecular Simulation,
Swinburne University of Technology,
Melbourne, Victoria 3122, Australia
}

\author{Peter J. Daivis}
\affiliation{
School of Applied Sciences, RMIT University, Melbourne, Victoria 3001, Australia
}

\author{Jesper S. Hansen}
\affiliation{
  DNRF Centre ``Glass and Time'', IMFUFA,
  Department of Sciences,
  Roskilde University, Postbox 260,
  DK-4000 Roskilde, Denmark
}

\begin{abstract}
The rheology and molecular structure of a model bitumen (Cooee bitumen)
under shear are investigated in 
the non-Newtonian regime using non-equilibrium molecular dynamics simulations.
The shear viscosity, normal stress differences and pressure of the bitumen
mixture are computed at different shear rates and different temperatures.
The model bitumen is shown to be a shear-thinning fluid at all temperatures.
In addition, the Cooee model is able to reproduce experimental results showing the formation
of nanoaggregates composed of stacks of flat aromatic molecules in bitumen. These
nanoaggregates are immersed in a solvent of saturated hydrocarbon molecules.
At a fixed temperature, the shear-shinning behavior is related to the
inter- and intramolecular alignment of the solvent molecules,
but also to the decrease of the average size of the nanoaggregates at high shear rates.
The variation of the viscosity with temperature at different shear rates is also related
to the size and relative composition of the nanoaggregates.
The slight anisotropy of the whole sample due to
the nanoaggregates is considered and quantified.
Finally, the position of bitumen mixtures in the broad literature of complex systems
such as colloidal suspensions, polymer solutions and associating polymer networks
is discussed.
\end{abstract}

\maketitle

\section{Introduction}

Bitumen is one of the by-products
of the refinery process of crude oil and a very viscous fluid.
Its main application is as a binder
in road pavement~[\onlinecite{shrp368}], because it combines many desirable properties:
it is cheap, it adheres quite well to mineral filler particles and has
suitable mechanical properties.
Among bitumen's mechanical properties, the shear viscosity is of special importance.
To guarantee a cohesive pavement,
the shear viscosity should not be too low, while high values are
often associated with a brittle behavior~[\onlinecite{shrp369, ec140}]
which facilitates the development of cracks and eventually leads to the deterioration
of the pavement.
Bitumen's viscosity also has a great effect on the exploitation of bitumen reservoirs
and transportation of bitumen.
This is why the experimental literature reporting viscosity
measurements of bitumen is so vast~[\onlinecite{johnson, shrp368, palade, ec140}].
Moreover, it is known that bitumen has a strong non-Newtonian behavior especially at low
temperatures~[\onlinecite{sybilski, ukwuoma, garcia, michalica}]. 
The non-Newtonian behavior is characterized by a strain rate dependent shear viscosity and
is believed to originate from the specific molecular structure of bitumen~[\onlinecite{dealyBook}],
containing in particular asphaltene nanoaggregates~[\onlinecite{mullins2011}].

Molecular dynamics (MD) is a good tool to study 
the molecular structure of bitumen and 
of the asphaltene nanoaggregates~[\onlinecite{murgich1996, pacheco2003, frigerio2011, ungerer2014}].
MD simulations can also quantify the mechanical properties
of the material. Indeed, there are a few numerical studies relating
the shear viscosity of bitumen in the limit of vanishing strain rate
to its molecular and supramolecular structure~[\onlinecite{zhang2007, us, aging, liGreenfield}].
However, to our knowledge, there has been no MD study relating the non-Newtonian
behavior of bitumen to its molecular structure, in particular to the nanoaggregate structure.
That is the first aim of this article.
A second aim is to compare the molecular structure of bitumen under shear to
that of other complex systems
such as colloidal suspensions~[\onlinecite{coussot}], polymer solutions~[\onlinecite{winkler}],
and associative polymer networks~[\onlinecite{vaccaro}]. The hope is then to
gain some insights on which structural characteristics of bitumen to focus
on to model its rheology.

In order to achieve these aims, we used non-equilibrium molecular dynamics simulations
with the SLLOD algorithm~[\onlinecite{evans, todd2007}]. This
algorithm is able to reproduce a shear Couette flow, even
far from equilibrium~[\onlinecite{evans}] and provide the value
of the shear rate dependent viscosity. 
The SLLOD algorithm has been used to simulate shear flow of complex systems such as 
hyperbranched
polymers~[\onlinecite{le}], associating polymer solutions~[\onlinecite{li}] and ionic liquids~[\onlinecite{van-Oanh}].
In this paper, we applied it to Cooee bitumen~[\onlinecite{cooee}],
which is a four-component bitumen model
describing a bitumen with generic mechanical properties~[\onlinecite{us}] and
molecular structure~[\onlinecite{aging, aggregate}].
Different temperatures were investigated in order to extrapolate the results
to ambient temperature.

The paper is organised as follows.
Section~\ref{sec:simuDetails} contains information about the 
model bitumen used and the SLLOD algorithm.
Results for the rheological properties
and the molecular structure of the sample are presented and
discussed in Secs.~\ref{sec:rheology} and~\ref{sec:molStruc}.
Finally, Sec.~\ref{sec:discussion}
contains a summary and a discussion.

\section{Simulation details}
\label{sec:simuDetails}

\subsection{Cooee bitumen}
\label{sec:cooeeModel}
\begin{figure}
  \scalebox{0.20}{\includegraphics{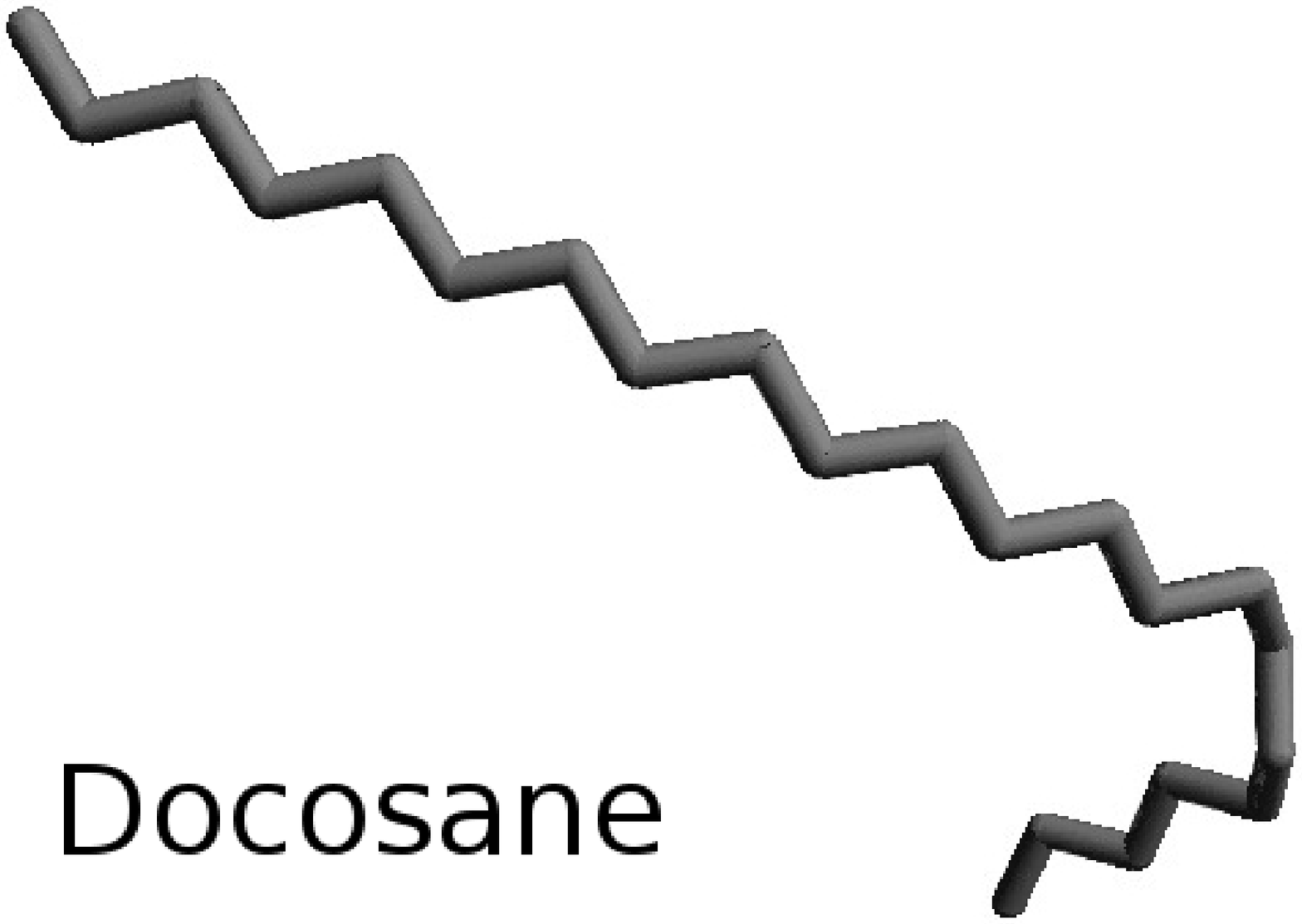}}
  \scalebox{0.20}{\includegraphics{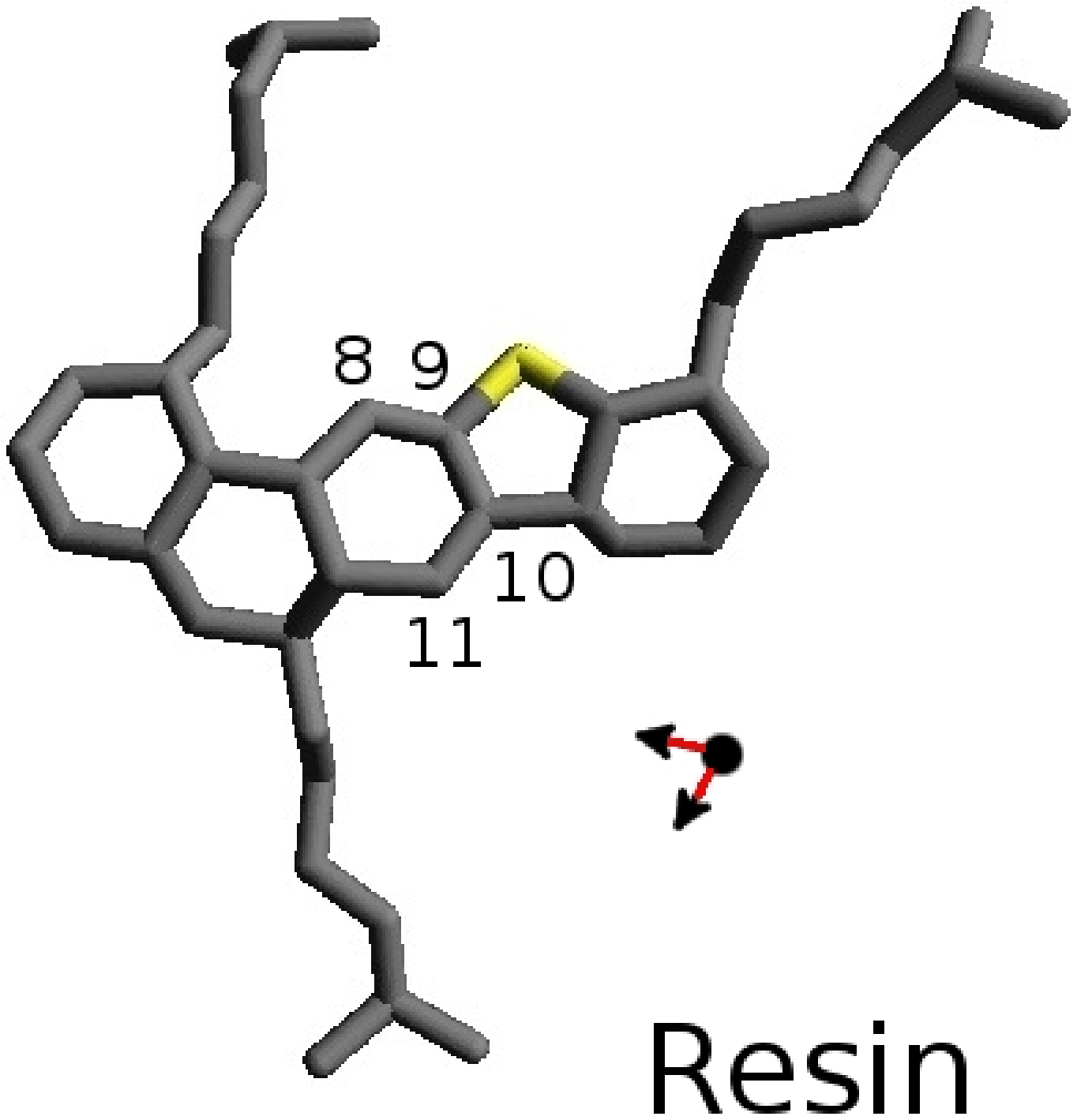}}\\
  \scalebox{0.25}{\includegraphics{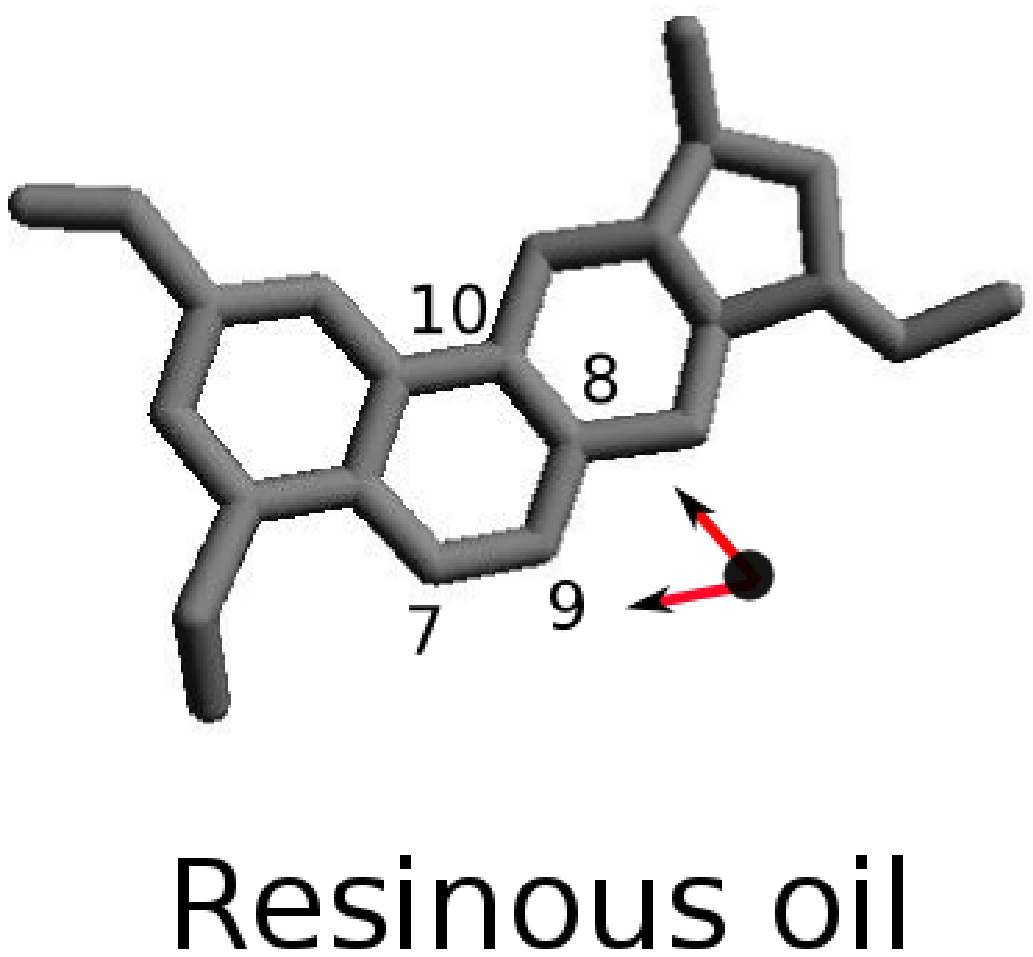}}
  \scalebox{0.25}{\includegraphics{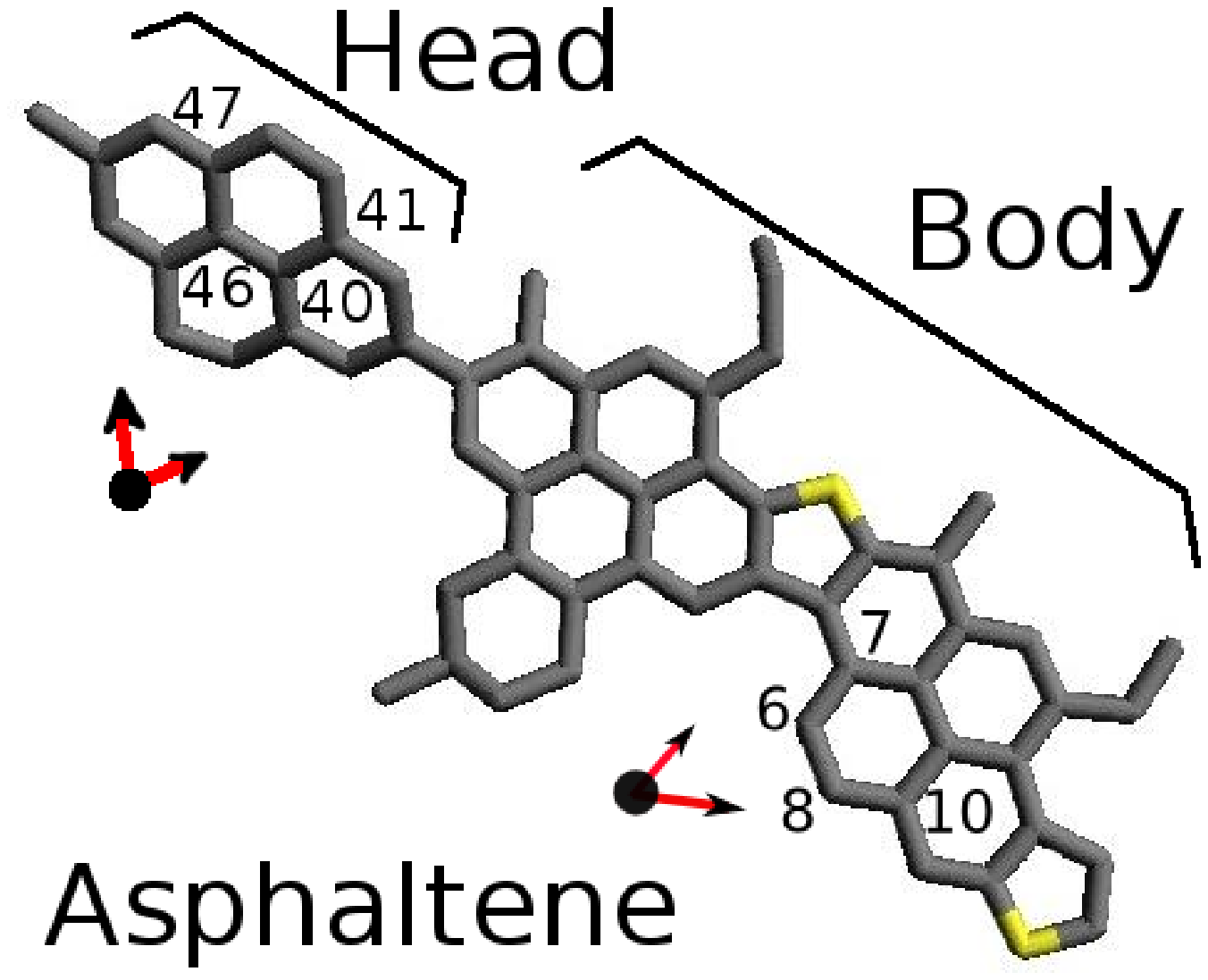}}
  \caption{\label{fig:molecule}
(Color online) Structure of the four molecules in the "Cooee bitumen" model. Grey edges represent
the carbon groups $CH_3$, $CH_2$ and $CH$ and yellow edges represent sulfur atoms.
The ``head'' and ``body'' of the asphaltene molecule are shown.
Numbers and arrows indicate bond-vectors used to quantify the nanoaggregate structure.
Reproduced from Ref.~[\onlinecite{aging}], (Copyright 2013 AIP).
}
\end{figure}

The chemical composition of bitumen is very complex~[\onlinecite{wiehe}].
However, the molecules in bitumen can be classified
into dissolution classes. The SARA classification~[\onlinecite{sara}] is one of the most common
classifications and it distinguishes between Saturated hydrocarbons, resinous oil molecules, called Aromatic
in the SARA scheme, Resin molecules and Asphaltene molecules. All of the last three molecule types are aromatic
and the molecular weight of asphaltene molecules is on average higher than that of resin molecules,
which is higher than the resinous oil molecular weight.
To obtain a simple model of a generic bitumen, one typical molecular structure for each class was chosen.
The molecular structures chosen are shown in Fig.~\ref{fig:molecule}. The reasons behind these specific choices
are given in Ref.~[\onlinecite{us}].
The main system studied in this paper contains
in weight fraction: $57.1$\%~of saturated hydrocarbons, $7.6$\%~of resinous oil molecules,
$13.8$\%~of resin molecules, and $21.5$\%~of asphaltene molecules.
The total number of united atom units in the system is~$15570$.
The methyl (CH$_3$), methylene (CH$_2$), and methine (CH) groups are represented by the
same united atom unit of molar mass~13.3~g$\cdot$mol$^{-1}$ and the sulfur atoms are represented
by a united atom unit with a different molar mass~32~g$\cdot$mol$^{-1}$.
The interaction potential between the united atom units contains four terms.
The first term corresponds to an intermolecular potential.
It is a Lennard-Jones potential
of the form $U_{LJ}(r) = 4\epsilon ((\sigma/r)^{12}- (\sigma/r)^{6})$
with parameters $\sigma = 3.75$~\AA $ $ and $\epsilon/k_B = 75.4$~K,
where $r$ is the interatomic distance and
$k_B$ the Boltzmann constant.
The method of shifted potential is used with a cutoff of $2.5$ $\sigma$.
The three other terms of the interaction potential describe 
intramolecular interactions. They control the bond length between two connected particles,
the angle between three consecutive particles, and the dihedral angle between four consecutive particles.
They are given by the following expression:
\begin{equation}
\begin{aligned}
  U_{\text{intra}}(\vec{r})&=  \frac{1}{2}\sum_{\text{\tiny{bonds}}}k_s(r_{ij}-l_{\text{\tiny{b}}})^2\\
  &+\frac{1}{2}\sum_{\text{\tiny{angles}}}k_{\theta}(\cos \theta - \cos
  \theta_0)^2
  +\sum_{\text{\tiny{dihedrals}}}\sum_{n=0}^5 c_n\cos^n \phi.
\end{aligned}
\label{eq:forcefield}
\end{equation}
The values of the parameters~$k_s$, $l_b$, $k_{\theta}$, $\theta_0$, and $c_n$ are listed in a previous work~[\onlinecite{us}].
The simulations were performed at four different temperatures: $377$~K,
$452$~K, $528$~K, and~$603$~K and at constant density.
For each temperature, the chosen density 
corresponds to an average equilibrium pressure equal to the atmospheric
pressure. The four densities chosen are:
$1007$~kg$\cdot$m$^{-3}$, $964$~kg$\cdot$m$^{-3}$, $894$~kg$\cdot$m$^{-3}$, and~$833$~kg$\cdot$m$^{-3}$.

\subsection{SLLOD algorithm}
\label{sec:sllod}
In this work, the molecular version of the SLLOD algorithm~[\onlinecite{evans, le}] is used
to obtain a shear flow of the bitumen mixture.
The equations of motion in that case are given by:
\begin{align}
\mathbf{\dot{r}}_{i \alpha} &=  \frac{\mathbf{p}_{i \alpha}}{m_{i \alpha}} + \mathbf{r}_i\cdot \boldsymbol{\nabla} \mathbf{u},\\
\mathbf{\dot{p}}_{i \alpha} &= \mathbf{F}_{i \alpha} - \frac{m_{i \alpha}}{M_i}\mathbf{p}_i\cdot \boldsymbol{\nabla} \mathbf{u} - \zeta \frac{m_{i \alpha}}{M_i} \mathbf{p}_i,
\end{align}
where $\mathbf{r}_{i \alpha}$ and $\mathbf{p}_{i \alpha}$ are the laboratory position
and peculiar momentum of particle~$\alpha$ in molecule~$i$, respectively 
(peculiar means that the corresponding velocities are relative to the streaming/flow velocity),
$\mathbf{F}_{i \alpha}$ is the force acting on particle~$\alpha$ in molecule~$i$,
$\nabla \mathbf{u}$ is the velocity gradient tensor, chosen here such that the $(x,y)$ component is
$\partial u_x/\partial y = \dot{\gamma}$ and all other components are zero,
$\mathbf{r}_i = \sum_{\alpha=1}^{N_i} m_{i \alpha} \mathbf{r}_{i \alpha}/M_i$
is the position of the center of mass of molecule~$i$, $M_i = \sum_{\alpha=1}^{N_i} m_{i \alpha}$
is the mass of molecule~$i$, $\mathbf{p}_i = \sum_{\alpha=1}^{N_i} \mathbf{p}_{i \alpha}$
is the momentum of the center of mass of molecule~$i$, $N_i$ is the number of particles in molecule~$i$,
and $\zeta$ is the Gaussian thermostat multiplier
given by:
\begin{equation}
\zeta = \frac{\sum_{i=1}^{N_m}\mathbf{F}_{i}\cdot\mathbf{p}_i/M_i - \dot{\gamma} \sum_{i=1}^{N_m} p_{ix}p_{iy}/M_i}{\sum_{i=1}^{N_m} \mathbf{p}_i^2/M_i},
\end{equation}
where $\mathbf{F}_{i} = \sum_{\alpha=1}^{N_i} \mathbf{F}_{i \alpha}$ is the force acting
on molecule~$i$ and $N_m$ the total number of molecules.
The expression for the thermostat multiplier $\zeta$
corresponds to the application of Gauss' principle of least constraint,
used to keep the molecular center of mass kinetic temperature constant.
The algorithm used to propagate the SLLOD equations of motion
is the operator splitting algorithm developed by Pan \textit{et al}~[\onlinecite{pan}] and adapted 
for molecular SLLOD equations of motion.
To our knowledge, it is the first time that the operator splitting algorithm has been adapted to molecular
systems. This implementation is described in appendix \ref{operatorsplitting}.
The MD graphical processing unit package RUMD~[\onlinecite{rumd}] was used to perform the calculation
in double precision.
The time step is equal to $\delta t = 0.86$~fs.
Each simulation was equilibrated for a period of~$1.7$~ns
and lasted thereafter $6.9$~ns. Depending on the strain rate and the temperature, $8$ or $16$ independent
initial configurations were used. 
The change in density from one initial configuration to the other is less than
$10^{-12}$~$\%$. 
The averages and standard errors shown in this paper
correspond to the averages and standard errors over these independent simulations.


\section{Rheology}
\label{sec:rheology}
Four main rheological properties can be computed
from the NEMD
simulations under shear. They are the shear viscosity~$\eta$, the first and second
normal stress coefficients~$\Psi_1$ and~$\Psi_2$, and the non-equilibrium
molecular pressure~$P$. They provide
information about the Newtonian and non-Newtonian behavior of the fluid.

These quantities are derived from the molecular stress tensor~$\boldsymbol{\sigma}$, given
by the Irving-Kirkwood formula~[\onlinecite{irving}]:
\begin{equation}
\label{eq:stress}
\boldsymbol{\sigma} = -\frac{1}{V}\biggl \langle \sum_{i=1}^{N_m} \frac{\mathbf{p}_i \mathbf{p}_i}{M_i} + \sum_{i=1}^{N_m} \sum_{j>i}^{N_m} \mathbf{r}_{ij} \mathbf{F}_{ij} \biggr \rangle,
\end{equation}
where $V$ is the volume of the system, $\langle \cdot \rangle$ denotes a time average
over the non-equilibrium steady state,
$\mathbf{r}_{ij} = \mathbf{r}_{i} - \mathbf{r}_{j}$ is the vector between
the centers of mass of molecules~$i$ and~$j$,
$\mathbf{F}_{ij} = \sum_{\alpha} \sum_{\beta} \mathbf{F}_{i\alpha j\beta}$
is the intermolecular force acting on molecule~$i$ due to molecule~$j$ and
$\mathbf{F}_{i\alpha j\beta}$ is the force acting on particle~$\alpha$ in
molecule $i$ due to particle~$\beta$ in molecule~$j$.

\subsection{Shear viscosity}
\label{sec:viscosity}

\begin{figure}
  \includegraphics[scale=0.4, angle=-90]{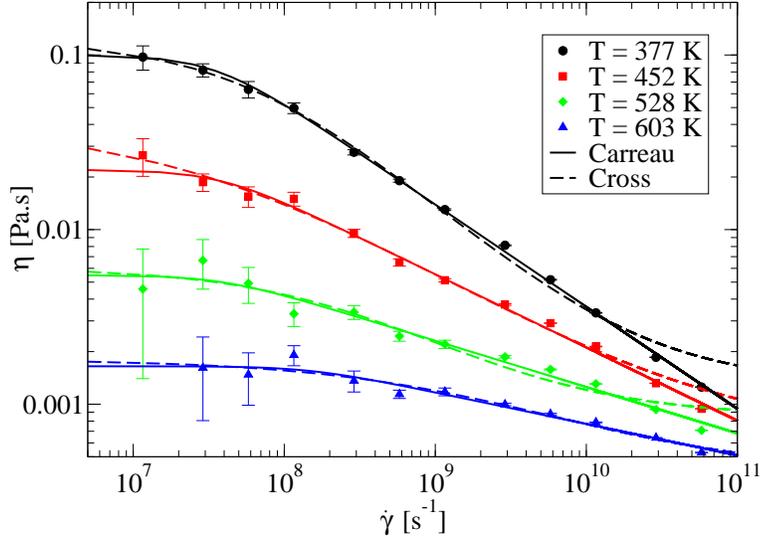}
  \caption{\label{fig:viscosityVsShearRate}
(Color online) Variation of shear viscosity~$\eta$ with the shear rate~$\dot{\gamma}$
for different temperatures.
The viscosity values corresponding to the four lowest shear rates
at temperature~$603$~K and~$528$~K  were averaged over 16 independent
simulations, whereas the other data points were averaged over 8 independent simulations.
The value of the viscosity at the lowest shear rate and at temperature~$603$~K is not
displayed because it is negative with an error bar larger than its absolute
value.
}
\end{figure}

Due to the shear geometry enforced in the simulation,
the shear viscosity is given by:
\begin{equation}
\eta = \frac{\sigma_{xy} + \sigma_{yx}}{2 \dot{\gamma}},
\end{equation}
where $\dot{\gamma}$ is the shear rate.
The variation of the viscosity~$\eta$ with the shear rate~$\dot{\gamma}$
is plotted in Fig.~\ref{fig:viscosityVsShearRate} for different temperatures.
At all temperatures the viscosity is decreasing with increasing shear rate
and the Cooee bitumen is consequently a shear-thinning fluid.
This is consistent with most experimental data on bitumen~[\onlinecite{sybilski, garcia, michalica}].
However, bitumen has a broad range of behavior with shear rate
and can also be shear-thickening~[\onlinecite{ukwuoma}].
Moreover, the higher the temperature, the higher the shear rate at which the fluid enters
the Newtonian regime in the Cooee model. Thus, the Newtonian plateau is clearly visible at
temperature~$603$~K and barely noticeable at temperature~$377$~K in the range
of shear rates accessible to NEMD simulations.
In other words, over the whole range of shear rates spanned, the
change in viscosity is larger at low temperatures than at high temperatures.
This is in agreement with the experimental literature~[\onlinecite{sybilski}],
stating a more pronounced non-Newtonian behavior of bitumen at low temperatures.

The zero shear rate viscosity~$\eta_0$ can be determined
using two empirical models to fit the data.
The models are known as the Carreau-Yasuda model~[\onlinecite{bird}] and the Cross model~[\onlinecite{farrington}].
The Carreau-Yasuda model suggests the following expression for the variation
of viscosity with shear rate: $\eta = \eta_0/[1+(\lambda \dot{\gamma})^2]^p$,
where $\lambda$ is a time constant
and $p$ a power law exponent.
The Cross model suggests a different empirical expression for the
variation of viscosity with shear rate:
$\eta = \eta_{\infty}+(\eta_0-\eta_{\infty})/[1+(K \dot{\gamma})^m]$,
where $\eta_{\infty}$ is the shear viscosity at infinite shear rate, $K$ the consistency index,
and $m$ a power law exponent.
The values of the parameters corresponding to the Carreau-Yasuda
and Cross models are given in Tables~\ref{tab:Carreau} and~\ref{tab:Cross}
for all temperatures. The values of the zero shear rate viscosity~$\eta_0$
should be taken with care for the two lowest temperatures as the Newtonian
regime is not clearly visible.

The values of these different parameters will now be discussed.
Both models obtain similar values for the viscosity at zero shear rate,
which gives some confidence in the quality of the data.
The experimental values for the zero shear rate viscosity at
a given temperature depend a lot on the precise origin and age of the
bitumen mixture studied. At around $377$~K the experimental
values for the viscosity can vary from $0.1$~Pa$\cdot$s~[\onlinecite{ukwuoma}]
to more than $100$~Pa$\cdot$s~[\onlinecite{sybilski, zhang2007}].
The value found in this work around~$0.1$~Pa$\cdot$s at~$377$~K is consequently
at the lowest end of the broad range obtained experimentally for different bitumen mixtures.

The values of the time constant~$\lambda$ in the Carreau-Yasuda model and of
the consistency index~$K$ in the Cross model are also quite close to each other.
They can be interpreted as the longest rotational relaxation time in the
system~[\onlinecite{van-Oanh}]. The rotational relaxation time of the asphaltene molecules
was estimated at temperature~$452$~K and density~$964$~kg$\cdot$m$^{-3}$
at equilibrium in an earlier work~[\onlinecite{aging}].
It was evaluated as the time needed for the rotational correlation function
to decay to $0.75$ and was found to be equal to $\tau_{\text{rot}} = 1.8\times 10^{-8}$~s.
This value is very close to the value of the time
constants $\lambda = 2.7\times 10^{-8}$~s and $K = 3.4\times10^{-8}$~s
found at the same temperature
and density in this work from the Carreau-Yasuda and Cross fits, respectively.

Finally, the Cross model does not seem to fit the data as well as the Carreau-Yasuda model
for high shear rates, despite that the former model was designed to improve the fit at high shear rates
by introducing a viscosity at infinite shear rate~$\eta_{\infty}$. The fact that the viscosity
keeps decreasing even at very high shear rates will be explained in terms
of molecular aggregation in Sec.~\ref{sec:aggregateSize}.

\begin{table}
\begin{center}
\begin{tabular}{c   c   c   c}
\hline
\hline
 $T$ [K]  & $\eta_0$ [Pa$\cdot$s]  & $\lambda$ [s] & $p$ \\
\hline
 $377$      & $0.10$              &  $2.9\times10^{-8}$      & $0.29$   \\
 $452$      & $0.022$             &  $2.7\times10^{-8}$      & $0.21$   \\
 $528$      & $0.0055$            &  $2.6\times10^{-8}$      & $0.13$   \\
 $603$      & $0.0016$            &  $6.7\times10^{-9}$      & $0.09$   \\
\hline
\hline
\end{tabular}
\end{center}
\caption{
\label{tab:Carreau}
Values of parameters~$\eta_0$, $\lambda$, and~$p$ from the Carreau-Yasuda
formula, for different temperatures~$T$. 
}
\end{table}

\begin{table}
\begin{center}
\begin{tabular}{c   c   c   c   c}
\hline
\hline
 $T$ [K]  & $\eta_0$ [Pa$\cdot$s]  & $K$ [s] & $\eta_{\infty}$ [Pa$\cdot$s] & $m$ \\
\hline
 $377$      & $0.13$             & $1.6\times10^{-8}$ &  $0.0013$                & $0.79$   \\
 $452$      & $0.040$            & $3.4\times10^{-8}$ &  $0.00064$               & $0.55$   \\
 $528$      & $0.0061$           & $4.2\times10^{-9}$ &  $0.00085$               & $0.70$   \\
 $603$      & $0.0018$           & $6.2\times10^{-10}$&  $0.00040$               & $0.56$   \\
\hline
\hline
\end{tabular}
\end{center}
\caption{
\label{tab:Cross}
Values of parameters~$\eta_0$, $K$, $\eta_{\infty}$, and~$m$ from the Cross
formula, for different temperatures~$T$.
}
\end{table}

\begin{figure}
  \includegraphics[scale=0.4, angle=-90]{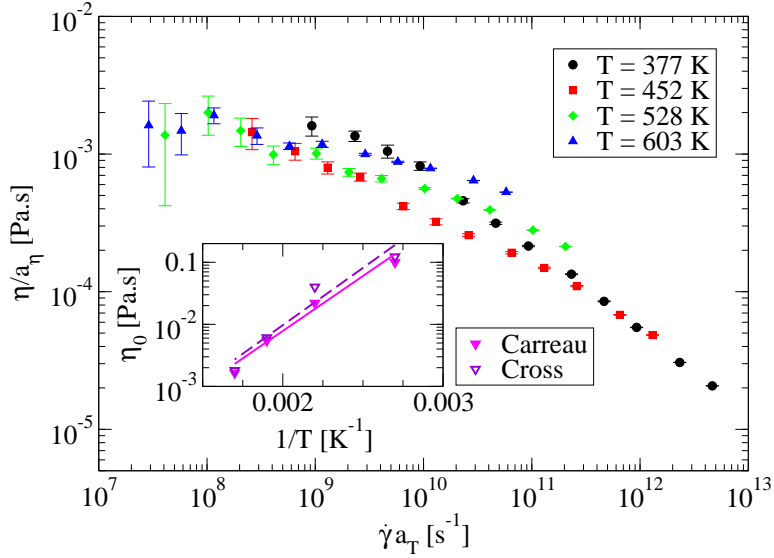}
  \caption{\label{fig:timeTempSuperposition}
(Color online) Variation of the reduced viscosity~$\eta/a_{\eta}$ with the
reduced shear rate~$\dot{\gamma} a_T$
for different temperatures. This is a test of the time-temperature
superposition principle.
Inset: (Color online) Variation of the zero shear viscosity~$\eta_0$,
evaluated from both the Carreau-Yasuda and the Cross models, with the
inverse temperature.
The solid lines are Arrhenius fits to the data.
}
\end{figure}

To extrapolate the data to experimentally relevant temperatures around~$300$~K, we
compare the variation of the zero shear rate viscosity with temperature to an Arrhenius behavior.
The Arrhenius equation of the form $\eta_0(T) = \eta^{\infty} \exp(E/(k_BT))$,
where $\eta^{\infty}$ is the viscosity at infinite temperature and
$E$ the activation energy, is usually valid for a range of temperatures
well above the glass transition~[\onlinecite{liu}]. Experimentally, the glass transition
of bitumen depends on its exact composition and is typically around~$250$~K~[\onlinecite{luIsacsson}],
which is well below the temperature range used in this work.
The Arrhenius equation has been shown to fit the variation of the experimental zero shear rate viscosity
with temperature quite well for some bitumen in a temperature range
going from $273$ to $363$~K~[\onlinecite{garcia}]. 
The variation with temperature of the zero shear rate viscosity obtained
with the Carreau-Yasuda and Cross fits is compared to an Arrhenius behavior
in the inset of Fig.~\ref{fig:timeTempSuperposition}.
Both sets of values follow an Arrhenius behavior. 
The parameters of the Arrhenius fit are found to be $E/k_B = 4073 $~K and $\eta^{\infty} = 2.27\times 10^{-6}$~Pa$\cdot$s
for the Carreau-Yasuda set of data and $E/k_B = 4245 $~K and $\eta^{\infty} = 2.00\times 10^{-6}$~Pa$\cdot$s.
Taking into account the data from the Carreau-Yasuda fit, the Arrhenius
equation predicts a value for the viscosity of~$2.5$~Pa$\cdot$s at~$293$~K.
Again, it is at the lowest end of the broad range of experimental values for the zero 
shear rate viscosity at this temperature. These values can range from $1$~Pa$\cdot$s~[\onlinecite{johnson}]
up to $10^4-10^6$~Pa$\cdot$s~[\onlinecite{zhang2007, garcia, palade, michalica}].

The fact that the viscosity of the Cooee bitumen is quite low
compared to experimental viscosities of bitumen at ambient pressure and temperature
can be related to the observation that the density of the Cooee bitumen is itself quite different
from experimental densities.
At a temperature of~$377$~K,
the density of the Cooee bitumen is $1007$~kg$\cdot$m$^{-3}$. It is to be compared
with the thorough experiments carried out on Athabasca bitumen~[\onlinecite{guan}]
which find a linear dependency of bitumen density on temperature at ambient pressure.
Extrapolating their results to $377$~K leads to a bitumen density around~$955$~kg$\cdot$m$^{-3}$,
which is significantly lower than the density of the Cooee bitumen at the same pressure and temperature.
This difference between the equation of state of our model and the equation of state
obtained experimentally could be due to the specific type of asphaltene molecules
that we chose. It has very short alkyl chains aside from the aromatic plane
compared to the resin molecule. We know from a previous work~[\onlinecite{aging}],
that adding resin molecules while removing asphaltene molecules from the mixture
has a tendency to lower the density at constant pressure. Thus, considering
asphaltene molecules with longer alkyl chains could lower the density of the model
and could also change the viscosity.
In the future, we plan to investigate this possible effect of the length of the alkyl chains
on the density and viscosity of the mixture by comparing two mixtures with
alkyl chains of different lengths. 

In an attempt to extrapolate not only the zero shear rate viscosity
but also the shear rate dependent viscosity to ambient temperature, we tested
the time-temperature superposition principle.
This principle has been shown to be generally valid experimentally~[\onlinecite{partal}]
and is also discussed in another numerical model of bitumen~[\onlinecite{masoori}].
The reduced viscosity and reduced shear rate are defined as $\eta/a_{\eta}$ and 
$\dot{\gamma} a_T$, respectively, where
\begin{align}
\label{eq:aT}
a_{\eta} &= \frac{\eta_0(T, \rho)}{\eta_0(T_{\text{ref}}, \rho_{\text{ref}})},\\
a_T &= a_{\eta} \frac{T_{\text{ref}} \rho_{\text{ref}}}{T\rho}.
\end{align}
The zero shear rate viscosity~$\eta_0$ is chosen from the Carreau-Yasuda fit,
the reference temperature~$T_{\text{ref}}$
and the reference density~$\rho_{\text{ref}}$ are chosen to be $603$~K
and~$833$~kg$\cdot$m$^{-3}$ respectively, because the Newtonian regime at this state point is the most clearly defined.
The procedure to obtain the reduced viscosity and reduced shear rate
is described in Ref.~[\onlinecite{bair}] and its theoretical foundation is discussed in
Ref.~[\onlinecite{bird}]. 
In particular, the time temperature superposition principle can be shown
to be valid if all the relaxation times in the system have the same temperature
dependence~[\onlinecite{vanGurp}].
Figure~\ref{fig:timeTempSuperposition} shows the variation of the reduced
viscosity~$\eta/a_{\eta}$
with the reduced shear rate~$\dot{\gamma} a_T$ for the Cooee bitumen.
The time temperature superposition principle
is not so well satisfied for this model bitumen.

\begin{figure}
  \includegraphics[scale=0.4, angle=-90]{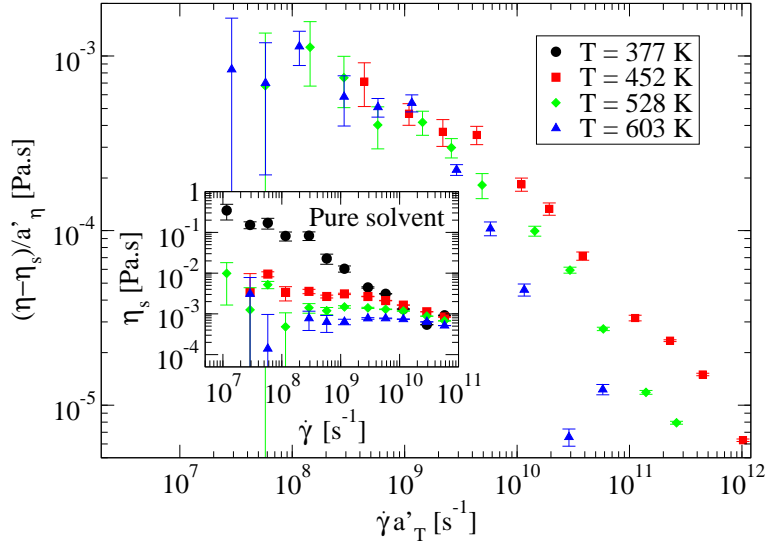}
  \caption{\label{fig:TTSsolvent}
(Color online) Variation of the reduced viscosity~$(\eta-\eta_s)/a'_{\eta}$ with the
reduced shear rate~$\dot{\gamma} a'_T$
at different temperatures.
Inset: Variation of the viscosity~$\eta_s$ with the shear rate~$\dot{\gamma}$
at different temperatures for the pure solvent.
}
\end{figure}

To explain this fact, we tried first to take into account the colloidal
nature of bitumen~[\onlinecite{mullins2011}].
Indeed, the Cooee bitumen
mixture can be seen as a suspension of branched nanoaggregates composed
of aromatic molecules and surrounded by a saturated hydrocarbon solvent~[\onlinecite{aggregate, aging}].
In the experimental literature,
bitumen mixtures are also sometimes treated as colloidal suspensions,
because the typical
size of a nanoaggregate is generally agreed to be around~$2$~nm~[\onlinecite{mullins2011}].
In the case of a dilute or semidilute suspension, the contribution
of the suspended objects to the total viscosity
can be well approximated by subtracting the pure solvent viscosity
to the total viscosity.
In the case of the Cooee bitumen studied here, the volume
fraction of aromatic molecules involved in the nanoaggregates
is $\phi = 0.42$, far from the dilute limit.
We tried nevertheless to subtract the viscosity of the pure solvent
at the same temperature and same density from the viscosity of the bitumen
mixture, to see if the time temperature superposition principle was better satisfied.
The reduced viscosity is now equal to~$(\eta-\eta_s)/a'_{\eta}$ and the reduced shear rate
to~$\dot{\gamma} a'_T$, where $\eta_s$ is the pure solvent viscosity at the same
temperature and density and where $a'_{\eta}$ and $ a'_T$ are given by:
\begin{align}
a'_{\eta} &= \frac{\eta_0(T, \rho) - \eta_s(T, \rho)}{\eta_0(T_{\text{ref}}, \rho_{\text{ref}}) - \eta_s(T_{\text{ref}}, \rho_{\text{ref}})},\\
a'_T & = a'_{\eta} \frac{T_{\text{ref}} \rho_{\text{ref}}}{T \rho}.
\end{align}
The reduced viscosity is plotted versus the reduced shear rate in Fig.~\ref{fig:TTSsolvent},
for the three highest temperatures. As expected, subtracting the solvent does not help
the time superposition principle to be satisfied.
Rather than a solvent effect, the failure
of the time superposition principle will be related
to the presence of at least two characteristic times related to the nanoaggregates
and with \textit{a priori}
different temperature dependences
in Sec.~\ref{sec:aggregateSize}.
The inset of Fig.~\ref{fig:TTSsolvent} displays the variation of the viscosity
of the pure solvent with shear rate
for different temperatures. It can be seen that at the lowest temperature, the pure solvent never
reaches the Newtonian plateau. It actually crystallizes. This crystallization is not seen
in the bitumen mixture. There could be two main reasons behind this: the melting temperature of the
mixture is lower than that of the pure solvent ; the bitumen mixture is blocked in a metastable liquid state
or a glassy state during the time spans accessible to MD whereas the pure
solvent is not.
Regardless of the possible crystallization of the solvent at the lowest temperature,
the fact that the
bitumen mixture is so far from the dilute limit indicates strongly that aggregate-aggregate
interactions are very important to understand the rheology of the mixture.

\subsection{Normal stress coefficients}
\label{sec:normalStressCoeff}

\begin{figure}
  \subfigure[]{\includegraphics[angle=-90, scale=0.25]{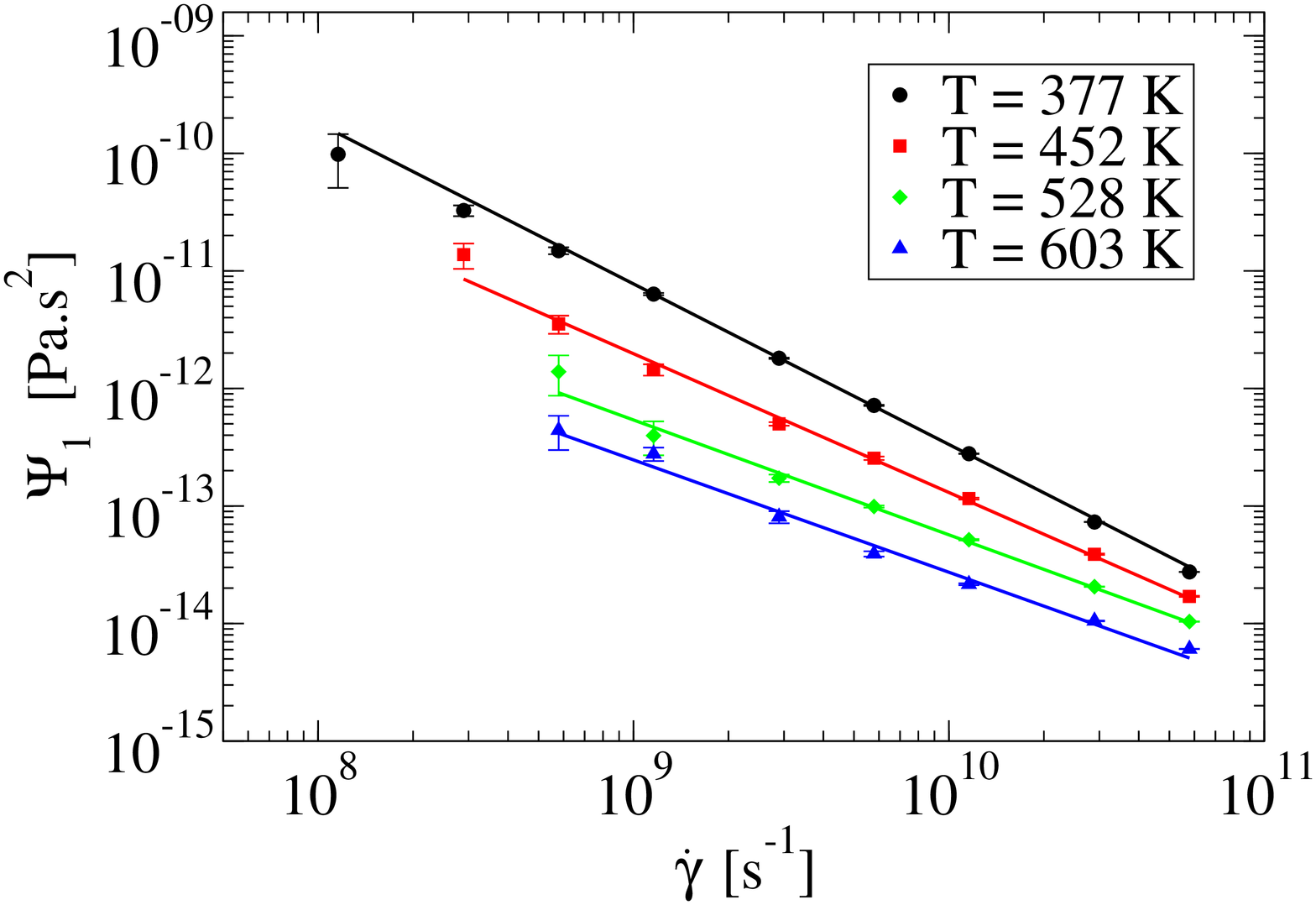}}
  \subfigure[]{\includegraphics[angle=-90, scale=0.25]{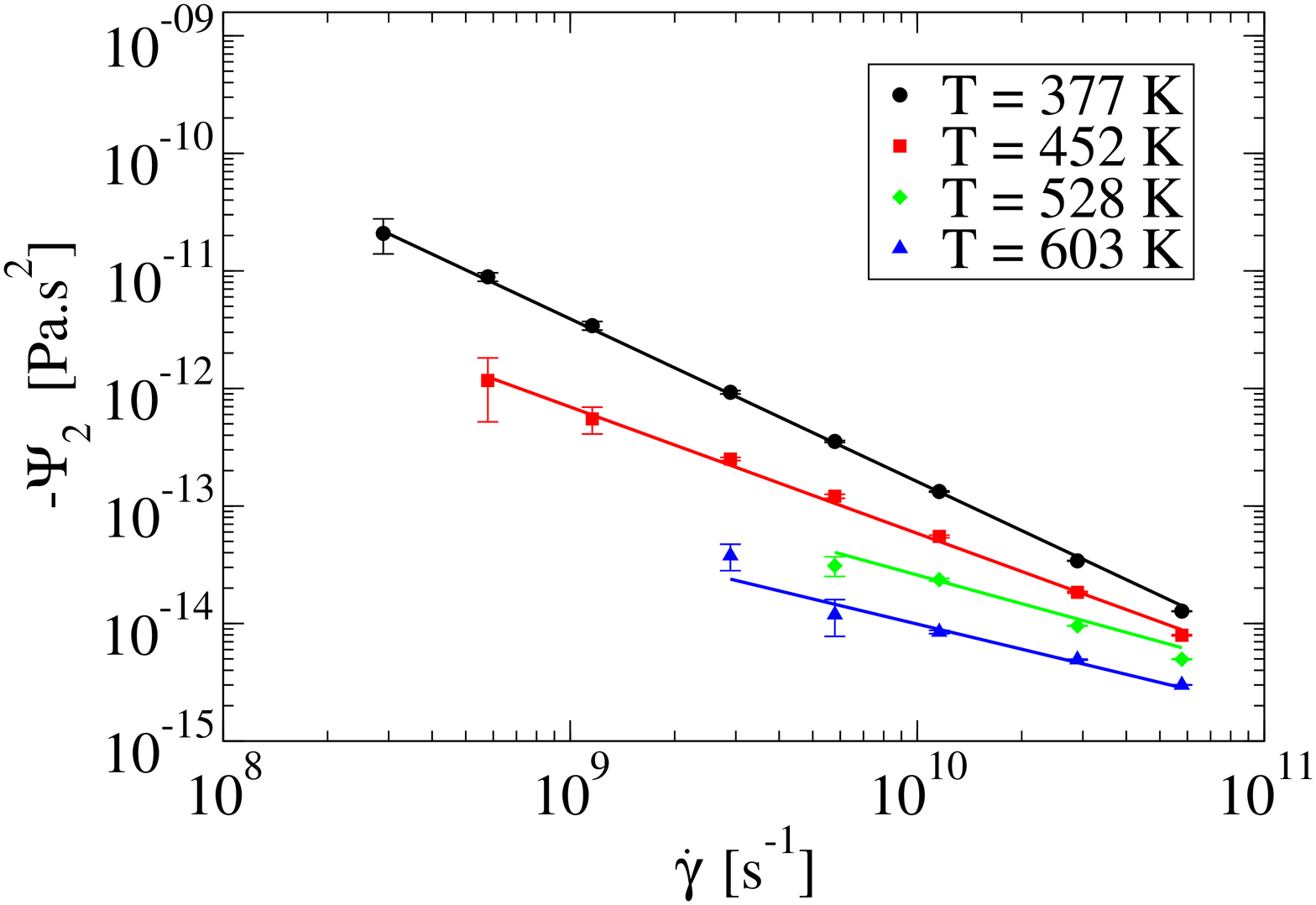}}
  \subfigure[]{\includegraphics[angle=-90, scale=0.25]{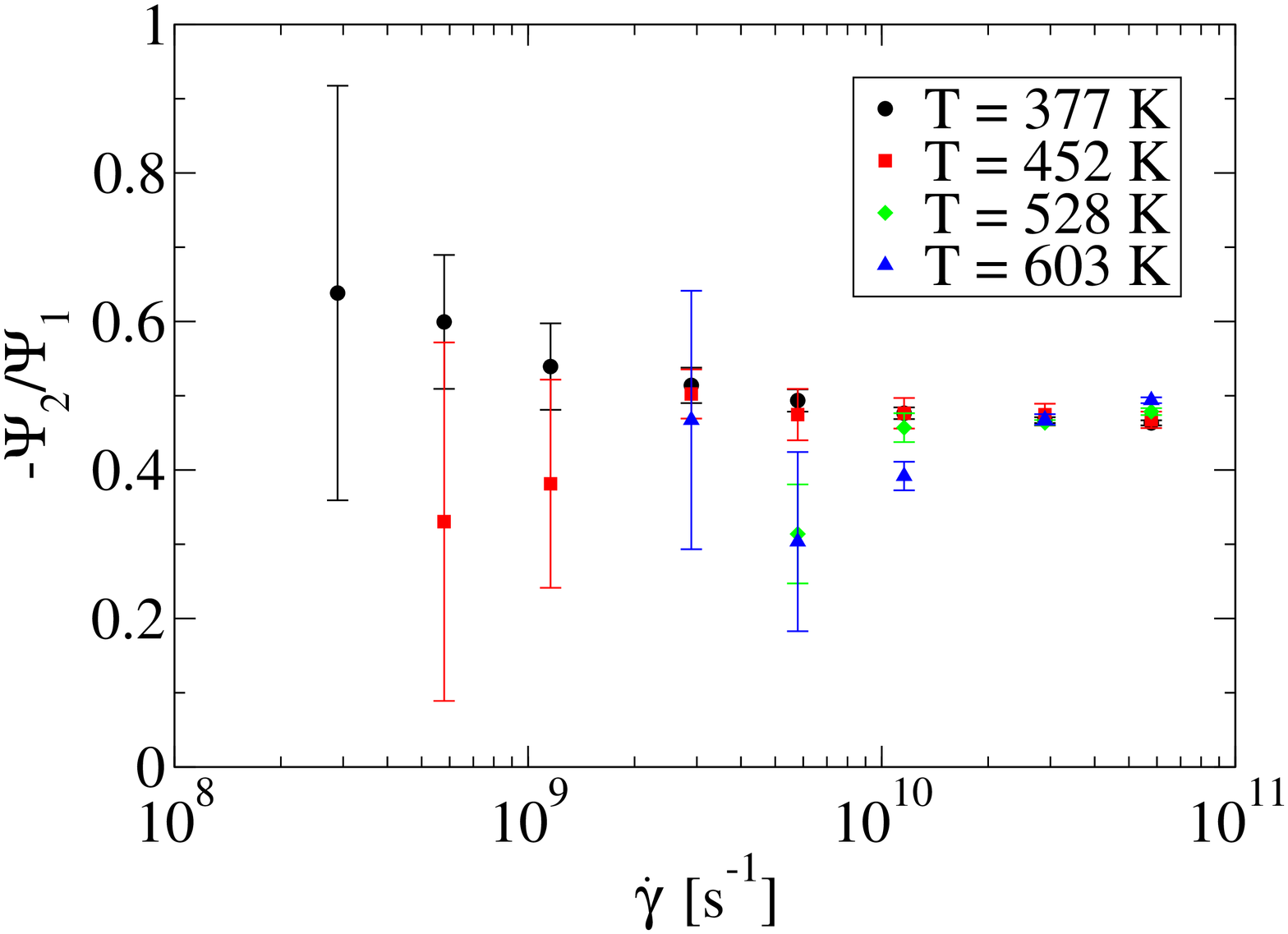}}
  \caption{\label{fig:normalStressCoeff}
(Color online) (a) and (b): Variation of the first~$\Psi_1$ and second~$\Psi_2$ normal stress coefficients with the shear rate~$\dot{\gamma}$
for different temperatures, respectively. In (b) $-\Psi_2$ is plotted versus shear rate.
The straight lines are power law fit to the data. Their exponents are given in Table~\ref{tab:exponentPsi}.
(c): Variation of the ratio~$-\Psi_2/\Psi_1$ with the shear rate~$\dot{\gamma}$ for the same temperatures.
}
\end{figure}

A fluid in the non-Newtonian regime
is also characterized by shear-rate dependent normal stress coefficients.
To first order in the shear rate, the two normal stress differences~$\sigma_{xx} - \sigma_{yy}$
and~$\sigma_{yy} - \sigma_{zz}$ are exactly zero because
a shear strain does not cause any change in the diagonal components
of the stress tensor.
Different models and experiments on polymers show that 
the first departures of the normal stress differences from zero occur
at second order in the strain rate~[\onlinecite{dealyBook}].
Therefore, the normal stress coefficients~$\Psi_1$ and~$\Psi_2$ are defined as:
$\Psi_1 = (\sigma_{xx} - \sigma_{yy})/\dot{\gamma}^2$
and $\Psi_2 = (\sigma_{yy} - \sigma_{zz})/\dot{\gamma}^2$.
Thus, at low shear rates, the normal stress coefficients~$\Psi_1$ and~$\Psi_2$
should plateau to a given value.
At larger shear rates, the normal stress coefficients are known to decrease
for polymers~[\onlinecite{dealyBook}].
The normal stress coefficients~$\Psi_1$ and~$\Psi_2$ obtained
from the NEMD simulations of Cooee bitumen are shown in Fig.~\ref{fig:normalStressCoeff}~(a) and (b),
respectively. Only the results having a standard error lower than their absolute value are displayed.
The first normal stress coefficient~$\Psi_1$ decreases with increasing shear rate~$\dot{\gamma}$,
showing again that bitumen has reached a non-Newtonian regime 
in the range of shear rates spanned in the simulations.
The second normal stress coefficient~$\Psi_2$ has been less studied,
but is usually negative in experiments on polymer fluids~[\onlinecite{le, bird}]. The simulations
presented in this paper also find a negative second normal stress coefficient~$\Psi_2$.
The absolute value of the second normal stress coefficient~$\Psi_2$ decreases with
increasing shear rate,
which is also a signature of the non-Newtonian regime.
The decrease of the two normal stress coefficients in the non-Newtonian regime
can be fitted to a power law~[\onlinecite{le}]:
$\Psi_1 \sim \dot{\gamma}^{-\alpha}$ and~$-\Psi_2 \sim \dot{\gamma}^{-\beta}$
for the coefficients~$\Psi_1$ and~$\Psi_2$, respectively.
The values of the exponents~$\alpha$ and~$\beta$ are displayed in Table~\ref{tab:exponentPsi}.
They are close to each other and decrease with increasing temperature, reaching
a value lower than~$1$ at the highest temperatures.

\begin{table}
\begin{center}
\begin{tabular}{c   c   c}
\hline
\hline
 $T$ [K]  & $\alpha$  & $\beta$  \\
\hline
 $377$      & $1.4$             &  $1.4$     \\
 $452$      & $1.2$             &  $1.1$     \\
 $528$      & $0.98$            &  $0.81$     \\
 $603$      & $0.96$            &  $0.71$     \\
\hline
\hline
\end{tabular}
\end{center}
\caption{
\label{tab:exponentPsi}
Values of exponents~$\alpha$ and~$\beta$ describing the power law decay of
the normal stress coefficients~$\Psi_1$ and~$\Psi_2$, respectively,
for different temperatures~$T$.
}
\end{table}

Finally, the normal stress ratio~$-\Psi_2/\Psi_1$ is plotted versus shear rate in Fig.~\ref{fig:normalStressCoeff}~(c).
This ratio is known to be equal to~$0.24$ for linear polymer melts and to~$0.3$
for branched polymer melts~[\onlinecite{dealyBook}].
In the case of bitumen, it is higher, around~$0.5$,
but does not seem to depend on either
shear rate or temperature. This value of the normal stress ratio~$-\Psi_2/\Psi_1$
will be related to an anisotropy effect in Sec.~\ref{sec:Sm}.

\subsection{Pressure}
\label{sec:normalPressure}

\begin{figure}
  \includegraphics[angle=-90, scale=0.4]{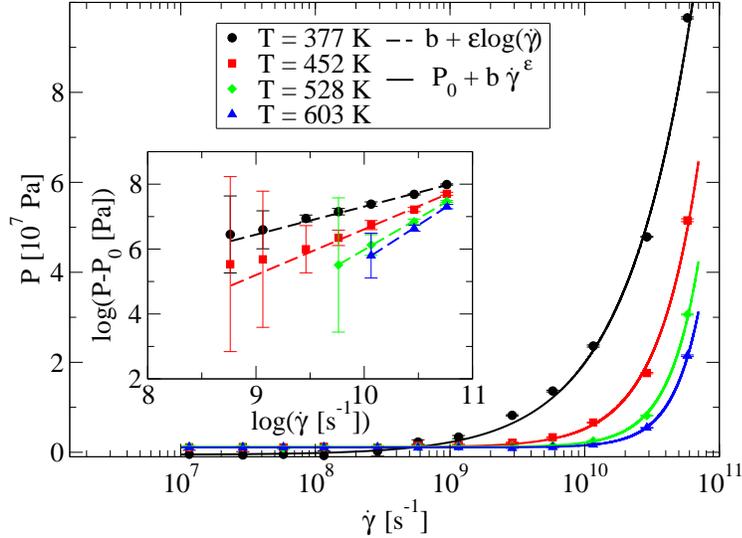}
  \caption{\label{fig:normalPressure}
(Color online)
Variation of the normal isotropic pressure~$P$ with the shear rate~$\dot{\gamma}$
for different temperatures.
Inset: Variation of $\log(P-P_0)$ with $\log(\dot{\gamma})$.
Only data with error bars lower than the value of $\log(P-P_0)$ are shown.
The dashed lines are straight line fits to this data. 
The solids lines in the main figure correspond to equation
$P = P_0 + b\dot{\gamma}^{\epsilon}$, where $b$ and $\epsilon$ were
determined by the straight line fits to $\log(P-P_0)$ versus $\log(\dot{\gamma})$.
The exponents $\epsilon$ are given in Table~\ref{tab:exponentPressure}.
}
\end{figure}

The pressure of the system is defined as minus a third of the trace of
the molecular stress tensor given in Eq.~\eqref{eq:stress}.
For simple fluids, the pressure~$P$ has been shown to have a power law
dependency on shear rate~[\onlinecite{ge, separdar}]: $P = P_0 + b\dot{\gamma}^{\epsilon}$,
where $P_0$ is the pressure at equilibrium.
For more complex fluids in particular for linear polymer melts, the pressure begins to decrease
with increasing shear rate before increasing with a power law behavior~[\onlinecite{bosko2004}].
The variation of the pressure with the shear rate is displayed in Fig.~\ref{fig:normalPressure}
for the Cooee bitumen model. The variation is found to be monotonic with the shear rate
for the range of temperatures explored. It can be fitted quite well with a power law. The
exponent of the power law is given for each temperature in Table~\ref{tab:exponentPressure}.
The observation that the pressure is monotonically increasing with the shear rate
instead of having a clear minimum
could be due to the fact that the molecules in bitumen
are not all linear. Moreover, it is known that the pressure
increases faster with shear rate for
high generation dendrimers than for linear polymers~[\onlinecite{bosko2004}].
This fast increase can mask the local minimum of the pressure.
The same mechanism could be at play in the bitumen mixture studied here.

Figure~\ref{fig:normalPressure} also shows that the density of the mixture
at each temperature is chosen so that the equilibrium pressure~$P_0$
is around the atmospheric pressure.

Finally, for shear rates lower than $3\times 10^8$~s$^{-1}$ and for all temperatures,
the pressure does not deviate significantly from its equilibrium value, enabling a direct
comparison with experimental results, usually obtained at constant pressure.

\begin{table}
\begin{center}
\begin{tabular}{c   c   c}
\hline
\hline
 $T$ [K]  & $\epsilon$   \\
\hline
 $377$      & $0.86$          \\
 $452$      & $1.4$           \\
 $528$      & $1.9$           \\
 $603$      & $2.2$           \\
\hline
\hline
\end{tabular}
\end{center}
\caption{
\label{tab:exponentPressure}
Values of the exponent~$\epsilon$ describing the power law increase of
the normal isotropic pressure~$P$
for different temperatures~$T$.
}
\end{table}

\section{Molecular structure}
\label{sec:molStruc}

\begin{figure}
  \includegraphics[scale=0.35, clip=true, trim=0cm 6cm 0cm 6cm]{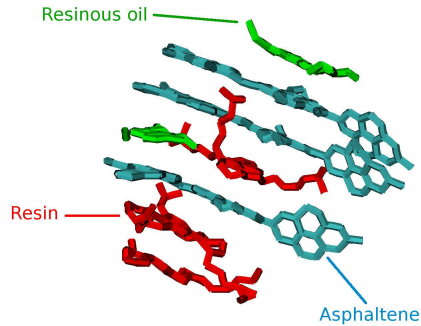}
  \caption{\label{fig:pictureAggregate}
(Color online) Snapshot of a linear nanoaggregate, obtained in equilibrium molecular dynamics. Asphaltene molecules
are in blue, resin molecules in red and resinous oil molecules in green.
Modified with permission from J. Chem. Phys. \textbf{141}, 144308 (2014).
}
\end{figure}

As mentioned in Sec.~\ref{sec:cooeeModel}, the Cooee bitumen model used in this paper
contains four molecule types. In order of increasing molecular weight, these types are:
docosane, resinous oil, resin and asphaltene. Of these, only the docosane molecules are not aromatic.
This composition resembles the SARA classification~[\onlinecite{sara}] and is able to reproduce
the characteristic supramolecular structure of bitumen in nanoaggregates~[\onlinecite{mullins2011}].
The nanoaggregates are composed of the aromatic molecules aligned with respect to each other. 
An example of a nanoaggregate seen in the simulations is shown in Fig.~\ref{fig:pictureAggregate}.
The docosane molecules can be seen as a solvent for the nanoaggregates.
The aim of this section is to quantify the variation of the nanoaggregate structure and
of the inter- and intramolecular structures
of each molecule type with shear rate and temperature as well as to compare it with the variation of the
rheological properties obtained in Sec.~\ref{sec:rheology}.

\subsection{Nanoaggregate size}
\label{sec:aggregateSize}

\begin{figure}
  \subfigure[]{\includegraphics[angle=-90, scale=0.25]{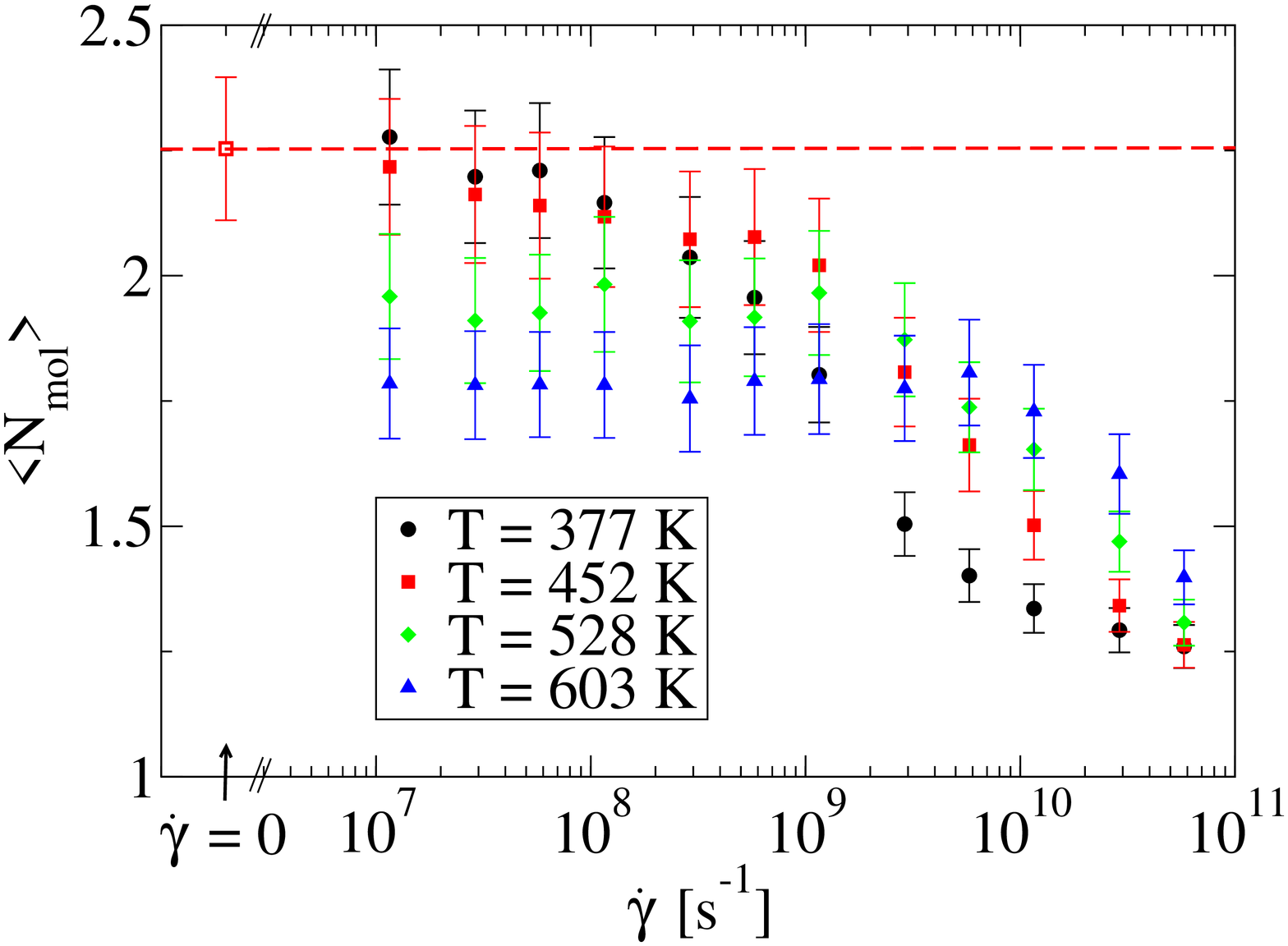}}
  \subfigure[]{\includegraphics[angle=-90, scale=0.25]{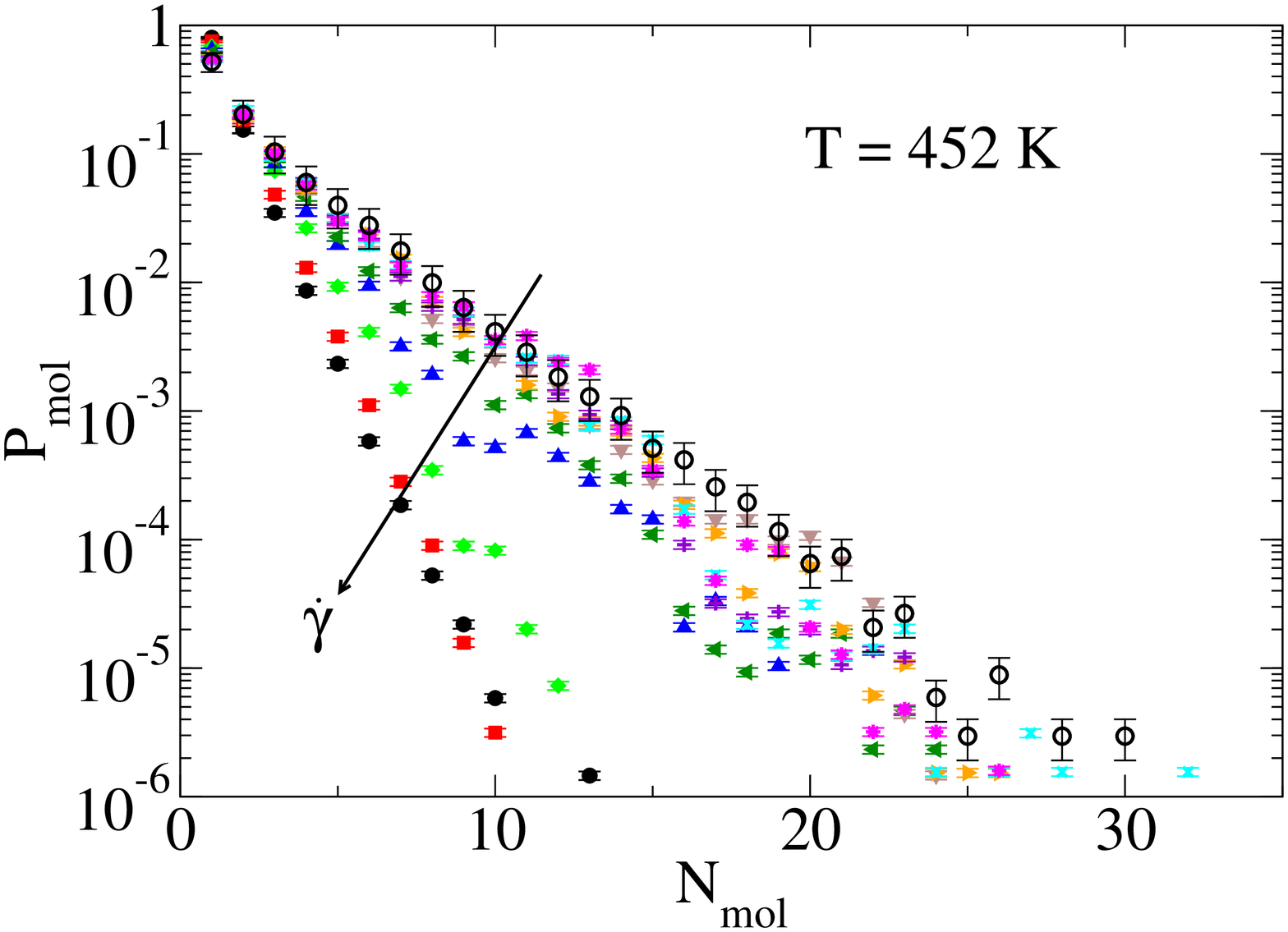}}
  \caption{\label{fig:AggregateSizeVsShearRate}
(Color online)
(a): Variation of the average number of aromatic molecules~$\langle N_{\text{mol}}\rangle$
in a linear nanoaggregate
with the shear rate~$\dot{\gamma}$ for different temperatures.
The open symbol and the dashed line represent the corresponding value at equilibrium and at temperature~$452$~K.
(b): Probability~$P_{\text{mol}}$ to have a linear nanoaggregate containing $N_{\text{mol}}$
molecules versus $N_{\text{mol}}$ at temperature~$452$~K and for different shear rates
from~$6\times 10^7$ to~$10 \times 10^{10}$~s$^{-1}$ (the two lowest shear rates are omitted
for the sake of clarity).
The open symbols correspond to the same result at equilibrium.
}
\end{figure}

The definition of a nanoaggregate from a molecular point of view is described in detail
in Refs.~[\onlinecite{aging, aggregate}] for equilibrium simulations. 
The nanoaggregates are composed of all the aromatic
molecules and are branched. The branched structure arises from the asphaltene molecules having
two flat parts, a head and a body, oriented in different directions.
Each part can align to other flat aromatic molecules creating a branched structure~[\onlinecite{aging}].
In this work, we focus on the linear segments of these branched structures,
because their size distribution has been studied in detail at equilibrium in an earlier work~[\onlinecite{aggregate}],
but the presence of the branched structure should be kept in mind.
The definition of a linear segment sums up to the following rule:
two aromatic molecules are nearest neighbors in the same nanoaggregate if they are ``well-aligned''
and ``close enough''. Only the alignment of aromatic molecules to an asphaltene
body and not to an asphaltene head is considered here, to extract purely linear structures.
All molecules connected by this rule are part of the same linear nanoaggregate.
The same rule was used in the NEMD simulations and
the average number of aromatic molecules in a linear nanoaggregate could be quantified for
different shear rates. The results are plotted in Fig.~\ref{fig:AggregateSizeVsShearRate}~(a)
for all temperatures.

\begin{figure}
  \subfigure[]{\includegraphics[angle=-90, scale=0.25]{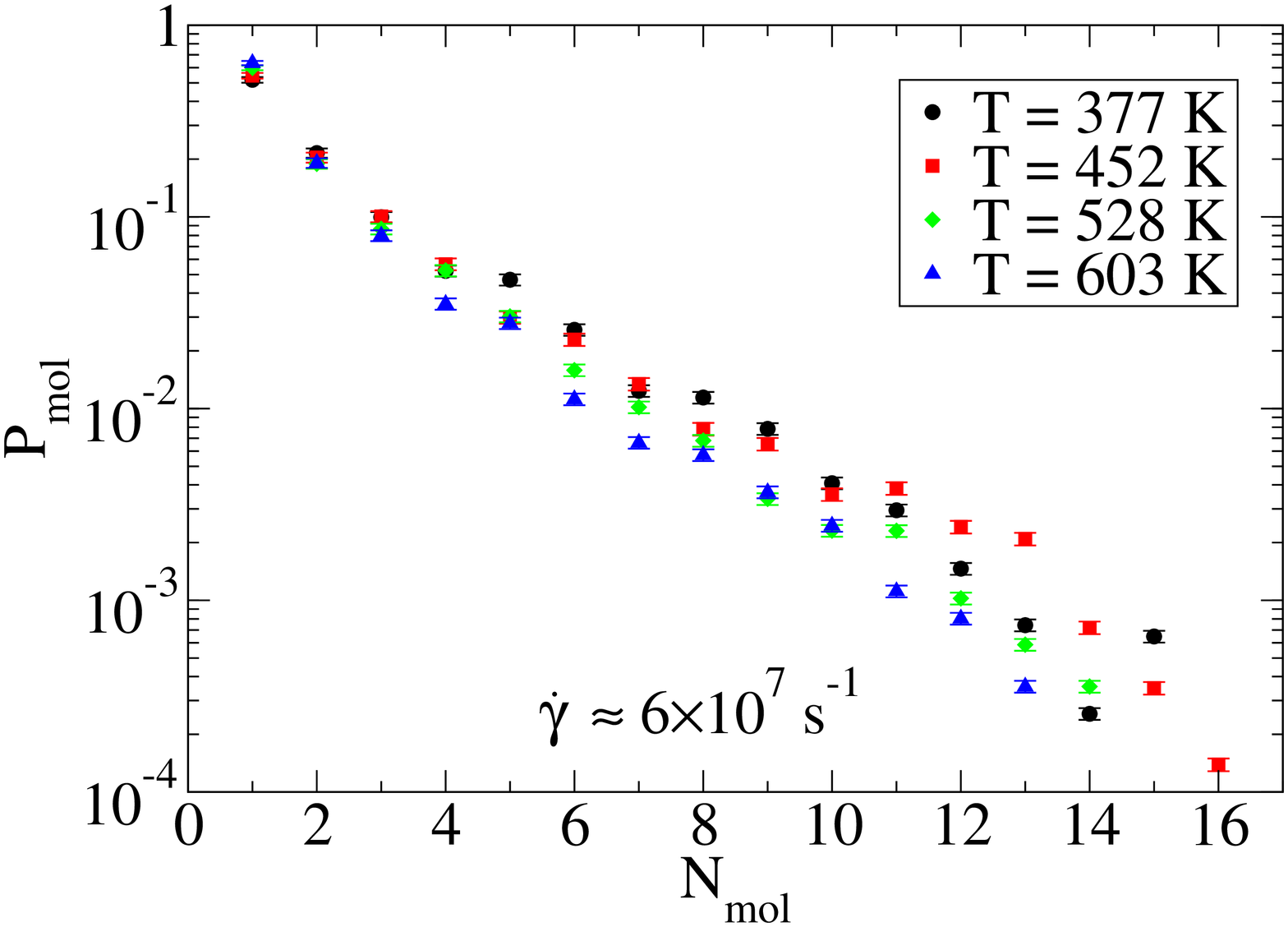}}
  \subfigure[]{\includegraphics[angle=-90, scale=0.25]{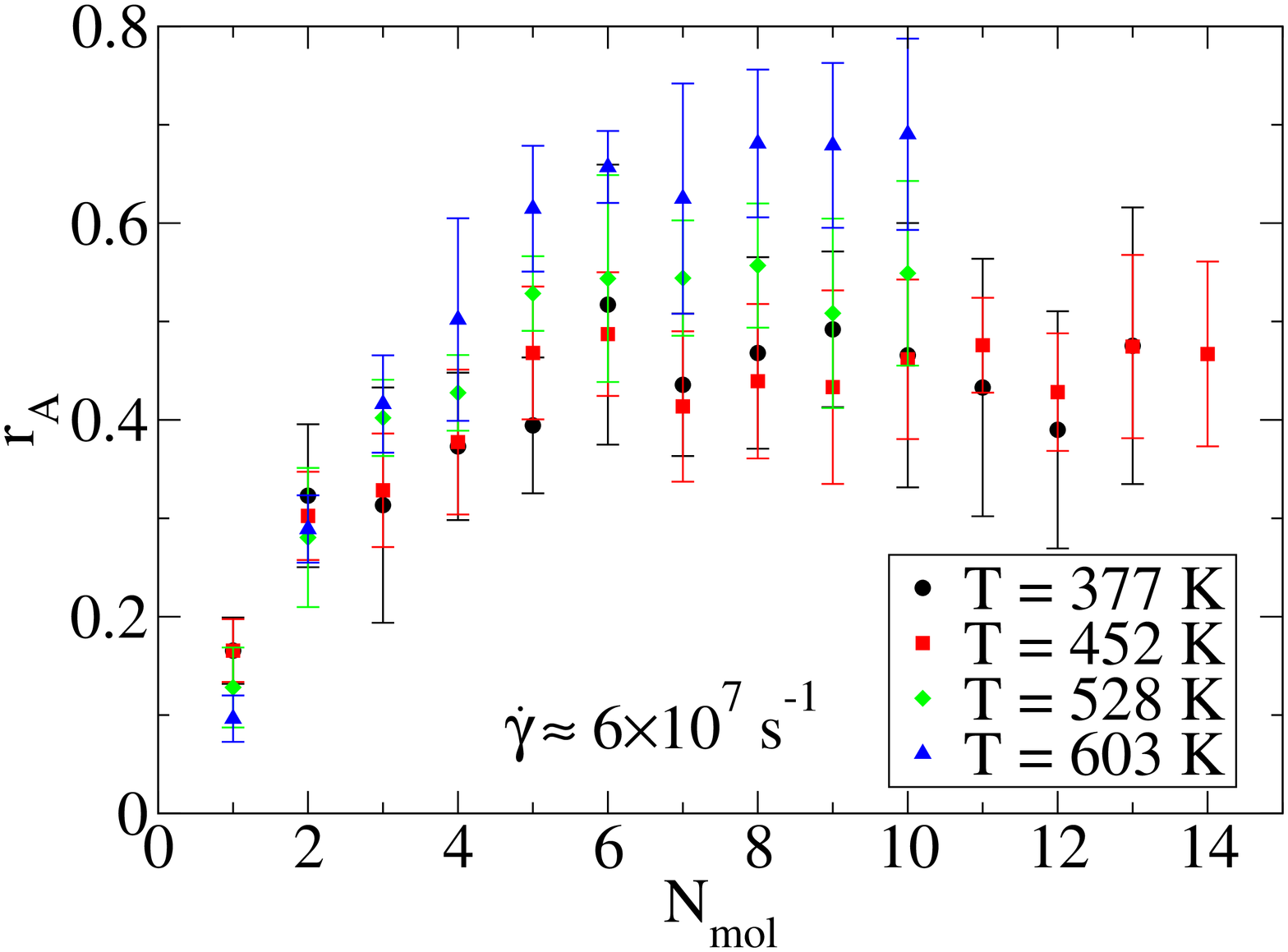}}
  \subfigure[]{\includegraphics[angle=-90, scale=0.25]{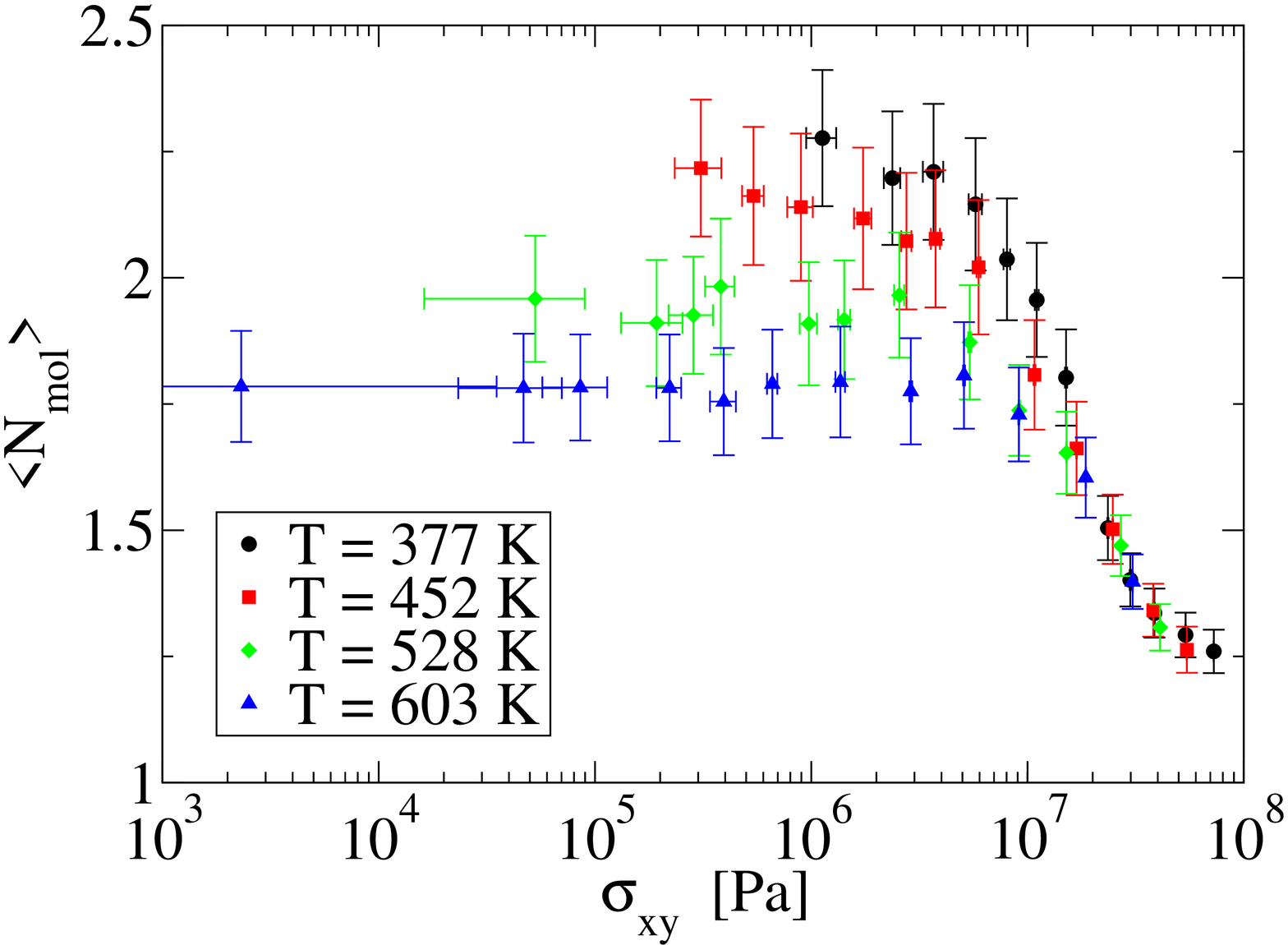}}
  \caption{\label{fig:AggregateSizeVsStress}
(Color online)
(a): Same as Fig.~\ref{fig:AggregateSizeVsShearRate}~(b)
for a given shear rate of~$6\times 10^7$~s$^{-1}$ and at different temperatures.
(b): Variation of the fraction~$r_A$ of asphaltene molecules in a nanoaggregate
with the number $N_{\text{mol}}$ of aromatic molecules in the same nanoaggregate,
at different temperatures and at
a fixed shear rate of~$6\times 10^7$~s$^{-1}$.
(c): Variation of the average number of aromatic molecules~$\langle N_{\text{mol}}\rangle$
in a linear nanoaggregate
with the stress~$\sigma_{xy}$ for different temperatures.
}
\end{figure}

The variation of the nanoaggregate size with the shear rate at a fixed temperature is first considered.
At low shear rates, the size of the nanoaggregates is close to their size at equilibrium
as is shown for temperature~$452$~K in Fig.~\ref{fig:AggregateSizeVsShearRate}~(a).
The full distribution
of the number of aromatic molecules in a linear nanoaggregate at low shear rates
is also very close to the same distribution at equilibrium
as can be seen in Fig.~\ref{fig:AggregateSizeVsShearRate}~(b).
At these low shear rates, the size distribution
has a biexponential shape, which we attributed, in
a previous work~[\onlinecite{aggregate}], to longer nanoaggregates
being less bent and containing more asphaltene molecules than smaller ones.
After a plateau at low shear rates, the nanoaggregate size decreases quite sharply with increasing
shear rate until it reaches a value close to~$1$,
indicating that nearly all nanoaggregates are reduced to single molecules.
In other words, the shear flow induces nanoaggregate rupture.
Figure~\ref{fig:AggregateSizeVsShearRate}~(b) shows that at the highest shear rate
and at temperature~$452$~K some longer nanoaggregates still remain but with a low probability.
Interestingly, figure~\ref{fig:AggregateSizeVsShearRate}~(b) also shows that at high
shear rates the size distribution is closer to a single exponential distribution.
However, no noticeable variation of the relative composition of the nanoaggregates
with their size was seen when the shear rate is increased (not shown).

The variation of the linear nanoaggregate size with temperature at a fixed shear rate
is now considered.
As can be seen in Fig.~\ref{fig:AggregateSizeVsShearRate}~(a),
the nanoaggregate size at low shear rates decreases with increasing temperature.
The size distribution of the linear nanoaggregates at a given low shear rate
appears biexponential at all temperatures as can be seen in
Fig.~\ref{fig:AggregateSizeVsStress}~(a).
The cause of the decrease of the nanoaggregate size with temperature
at low shear rates can be further investigated by looking at the relative composition
of the nanoaggregates for different temperatures.
The fraction~$r_A$ of asphaltene molecule in a nanoaggregate is defined
as the number of asphaltene molecules in this nanoaggregate
divided by the total number of aromatic molecules in the same nanoaggregate.
Figure~\ref{fig:AggregateSizeVsStress}~(b) shows the variation of the fraction~$r_A$
of asphaltene molecules with the nanoaggregate size for different temperatures
and at a low shear rate of~$6\times 10^7$~s$^{-1}$.
For linear nanoaggregates with a number of molecules larger than~$3$,
the fraction~$r_A$ of asphaltene molecules in the nanoaggregates
increases with increasing temperature.
Concurrently, the fraction
of resin and resinous oil molecules in the nanoaggregates tends to
decrease with increasing temperature in long nanoaggregates.
It means that the overall decrease of the nanoaggregate size
with temperature is mainly due to resin and resinous oil molecules
being excluded from the aggregates at large temperatures.
The asphaltene molecules stay aggregated even at large temperatures.
It is probably an effect of the cohesive
energy being higher in absolute value between two asphaltene molecules
than between two resin or two resinous oil molecules.
This is itself due 
the difference in the sizes of the aromatic structures. Asphaltene
molecules have a larger aromatic structure than
resin and resinous oil molecules, thus creating a larger $\pi$-stacking
interaction between the aromatic planes of two asphaltene molecules
than between two resin or resinous oil molecules.
At high shear rates in Fig.~\ref{fig:AggregateSizeVsShearRate}~(a),
it seems on the contrary that the nanoaggregate size
increases with increasing temperature.
It is due to the fact that the shear rate at which the nanoaggregates 
begin to break up
is not the same at all temperatures. 
The higher the temperature, the lower the shear rate needed to
break up the nanoaggregates.
This apparent crossover between the curves giving the nanoaggregate size
versus the shear rate at different temperatures
can be removed if the average nanoaggregate size is plotted versus the shear
stress~$\sigma_{xy}$ rather than the shear rate~$\dot{\gamma}$.
This can be seen in Fig.~\ref{fig:AggregateSizeVsStress}~(c).
This figure also shows that above a certain shear stress, all curves collapse.
Figure~\ref{fig:AggregateSizeVsStress}~(c) can be understood if one assumes that
the size of the nanoaggregates depends on two factors:
the thermal energy and the work done by the external shear force applied
to the system. The thermal energy is directly proportional to temperature
and the work done by the shear force can be estimated from the product
of the shear stress~$\sigma_{xy}$
and the characteristic volume of an aromatic molecule
in a nanoaggregate. 
At low shear stresses, the work done by the shear force is negligible
with respect to the thermal energy and the nanoaggregate size is determined by
thermal energy alone. The nanoaggregates have then their equilibrium size.
At high shear stresses, on the contrary, the thermal energy
is negligible with respect to the work done by the shear force and the nanoaggregate
size is determined by the shear stress~$\sigma_{xy}$ alone. 
The curves at different temperatures collapse at high shear stresses.
At intermediate
values of the shear stress, both factors play a role in the nanoaggregate size. 
In this intermediate regime, an increase in shear stress at constant temperature
results in a monotonic decrease of the nanoaggregate size and an increase in temperature
at constant shear stress results in a monotonic
decrease of the nanoaggregate size.

The connection between the rupture of the nanoaggregates
around a given shear rate and the variation of the viscosity with
shear rate and temperature quantified
in Sec.~\ref{sec:rheology} is now discussed.
The overall decrease of the nanoaggregate size with increasing temperature at low shear rates
contributes to the decrease of the viscosity observed with increasing
temperature in Fig.~\ref{fig:viscosityVsShearRate}.

The rupture of the nanoaggregates at high shear rates is consistent
with a decrease in the viscosity at these same shear rates instead of a
plateauing around the viscosity at infinite shear rate~$\eta_{\infty}$,
predicted by the Cross model. The Carreau-Yasuda model does not postulate the existence
of a constant viscosity at infinite shear rate. It could be why it better fits the data
than the Cross model, as was found in Sec.~\ref{sec:rheology}.

Moreover, the important change in the nanoaggregate size upon shearing could explain
why the time-temperature superposition principle is not
satisfied. Indeed, the time temperature superposition typically
works when all relaxation times have the same temperature
dependence~[\onlinecite{vanGurp}]. In the case of the Cooee bitumen model,
at least two main long relaxation times can be identified:
the rotational relaxation time of the asphaltene molecules,
close to the characteristic time~$\lambda$ of the Carreau-Yasuda fit,
and the inverse shear rate at which the nanoaggregates rupture.
They have \textit{a priori} different temperature dependences, which
could explain why the time temperature superposition principle is not satisfied.

\subsection{Molecular alignment}
\label{sec:molecularAlignment}

The variation of the size of the nanoaggregates, a supramolecular structure specific to bitumen, has been addressed
in Sec.~\ref{sec:aggregateSize}. This section (\ref{sec:molecularAlignment}) and the next (\ref{sec:intramol})
are devoted to more usual quantifications of the variation of the inter- and intramolecular structures
with the shear rate and their link to the shear-thinning behavior reported in Sec.~\ref{sec:rheology}.
This section focuses on 
the alignment of the molecules in the flow direction, also known as form birefringence.

Form birefringence is
a departure from the isotropic equilibrium state of the fluid due to the applied velocity gradient.
The anisotropy arises in this case from the alignment of
the molecules in the flow direction~[\onlinecite{doi}].
The form birefringence is quantified using a
molecular order tensor~$\mathbf{S}_m$, defined for a given molecule type as:
\begin{equation}
\label{eq:Sm}
\mathbf{S}_m = \frac{1}{N_m}\sum_{i=1}^{N_m} \Bigl \langle \mathbf{u}_i\otimes\mathbf{u}_i -\frac{1}{3}\mathbf{I}\Bigr \rangle,
\end{equation}
where $N_m$ is the total number of molecules of the considered type,
$\mathbf{u}_i$ is the unit vector denoting the principal direction of molecule $i$ of the given type,
$\mathbf{I}$ is the identity tensor, and $\langle \cdot \rangle$ denotes a time average
in the non-equilibrium steady state.
The principal direction~$\mathbf{u}_i$ of molecule~$i$ is the unit eigenvector corresponding
to the largest eigenvalue of the gyration tensor~$\mathbf{R}_g^2$, defined as:
\begin{equation}
\label{eq:tensorOfGyration}
\mathbf{R}_g^2 = \frac{1}{2N_i} \sum_{\alpha=1}^{N_i}\sum_{\beta=1}^{N_i} (\mathbf{r}_{\alpha}-\mathbf{r}_{\beta})\otimes (\mathbf{r}_{\alpha}-\mathbf{r}_{\beta}),
\end{equation}
where $N_i$ is the number of atoms in molecule~$i$ and $\mathbf{r}_{\alpha}$ is the position
of atom~$\alpha$ in molecule~$i$.
Once the molecular order tensor~$\mathbf{S}_m$ is determined for each molecule type, two
quantities derived from it are of interest: the molecular order parameter~$S_m$
and the molecular alignment angle~$\chi_m$.
The molecular order parameter~$S_m$ is
defined as~$3/2$ times the largest eigenvalue of the molecular order tensor.
It is equal to~$0$ when the molecules are 
randomly oriented and to~$1$ when all molecules are perfectly aligned.
The molecular alignment angle~$\chi_m$ is equal to arctan$(e_y/e_x)$, where
$\mathbf{e} = (e_x, e_y, e_z)$ is the eigenvector corresponding to the largest eigenvalue of the
molecular order tensor~$\mathbf{S}_m$.
The molecular alignment angle~$\chi_m$ represents the angle between the
average molecular orientation~$\mathbf{e}$ and the shear direction ($x$-direction in this case)
in the $xy$-plane.

\subsubsection{Molecular order parameter}
\label{sec:Sm}
\begin{figure}
  \subfigure[]{\includegraphics[angle=-90, scale=0.25]{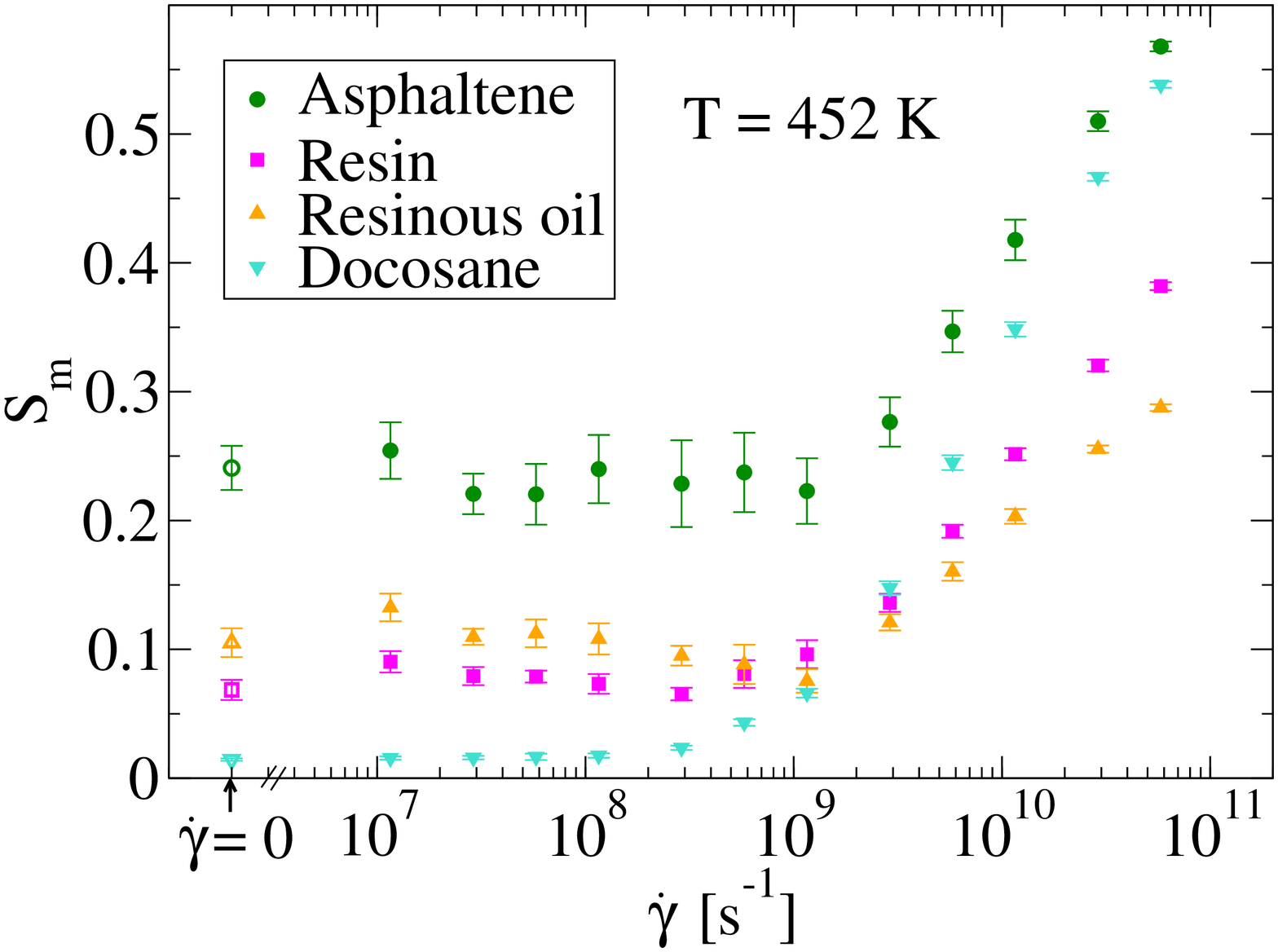}}
  \subfigure[]{\includegraphics[angle=-90, scale=0.25]{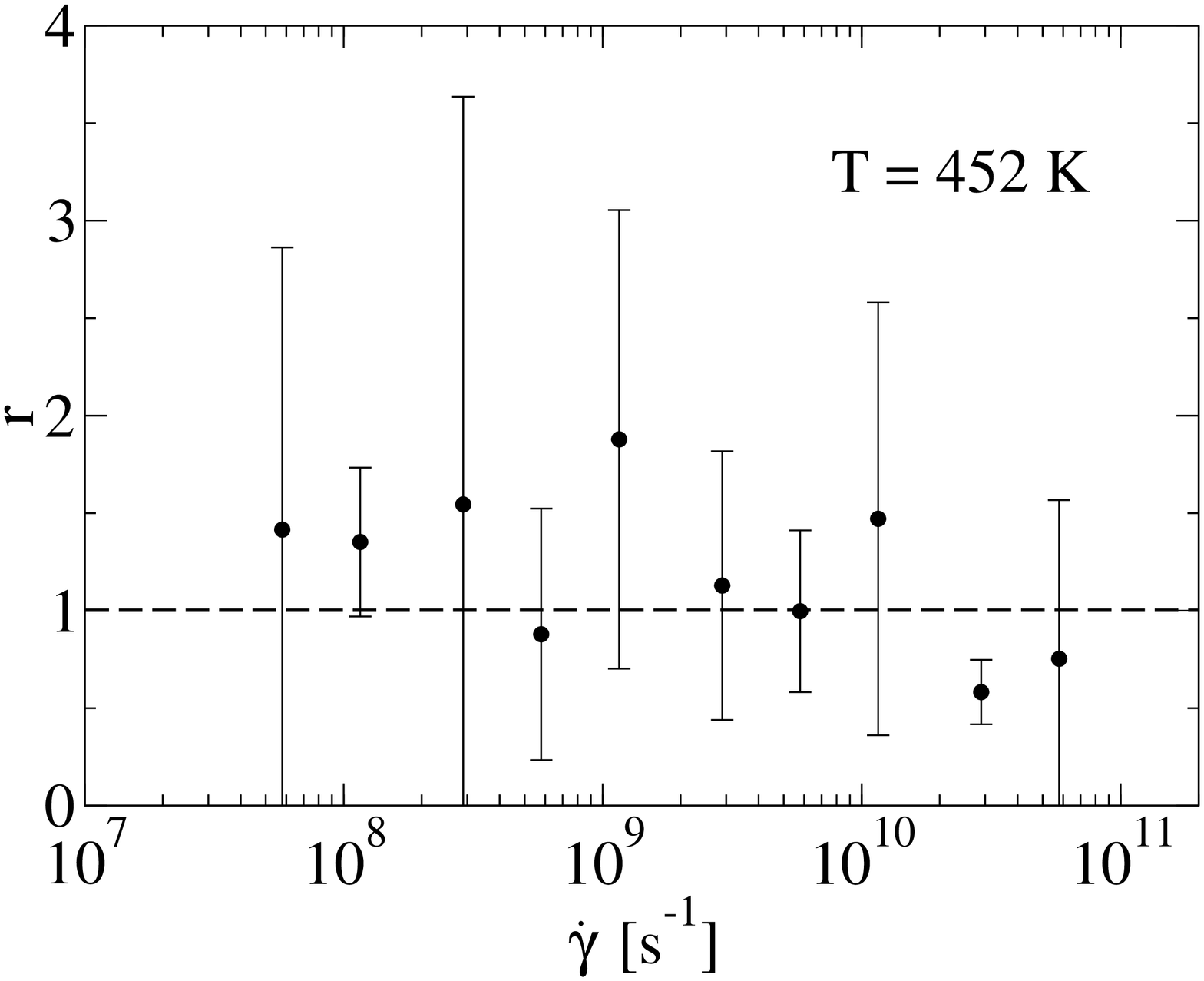}}\\
  
  \caption{\label{fig:molecularOrderParamVsShearRate}
(Color online)
(a):
 Variation of the molecular order parameter~$S_m$ with shear rate~$\dot{\gamma}$
for different molecule types and for a temperature of~$452$~K.
The open symbols represent the corresponding values at equilibrium.
(b): Variation of the ratio~$r = s_{\text{rot}}/s_{\text{conf}}$ comparing the standard deviation
due to the anisotropy of the sample to the standard deviation due to different
samples with the shear rate~$\dot{\gamma}$ at a temperature of~$452$~K.
The error bars are calculated for each type of standard deviations as
the absolute value of the difference between the standard deviation corresponding to half of the
data set and the standard deviation corresponding to the other half of the data set.
The relative error on the ratio~$r$ is then calculated as the sum of the
relative errors on~$s_{\text{rot}}$ and~$s_{\text{conf}}$.
The dashed line has equation~$r = 1$ and corresponds
to the anisotropy of the sample being the main source of variation between one
initial configuration and another.
}
\end{figure}

The variation of the molecular order parameter~$S_m$ with the shear rate~$\dot{\gamma}$
is displayed in Fig.~\ref{fig:molecularOrderParamVsShearRate}~(a) for a temperature of~$452$~K
and for the different molecule types.
The curves are quite different from one molecule type to another. 
For docosane molecules, the molecular order parameter~$S_m$ tends to zero
in the limit of vanishing shear rate. This behavior is expected, since
the docosane molecules are isotropically distributed at equilibrium. 
The molecular order parameter for docosane increases monotonically
to reach the value~$0.55$ at the highest shear rate simulated,
showing an alignment of docosane molecules in the shear direction.
This is consistent with the shear thinning behavior globally observed for
Cooee bitumen~[\onlinecite{forster}].
For asphaltene, resin and resinous oil molecules the curve is qualitatively different.
At low shear rates, until a critical shear rate of approximately
$\dot{\gamma}_c = 2\times 10^9$~s$^{-1}$ at temperature~$452$~K is reached, the
molecular order parameter of the three aromatic types is roughly constant and non-zero. Above this critical
shear rate, the molecular order parameter begins to increase with the shear rate.
The small finite value reached by the molecular order parameter at low shear rates
is equal to the one obtained
from equilibrium molecular dynamics simulations, as can also be seen in
Fig.~\ref{fig:molecularOrderParamVsShearRate}.
This value is higher for asphaltene molecules than for resinous oil molecules
and for resin molecules. 
This unusual behavior of the molecular order parameter is easily explained
if the presence of nanoaggregates at low shear rates is considered.
Indeed, the long nanoaggregates tend to align with respect to each other,
due to steric hindrance, creating
a slight anisotropy in the system. This anisotropy is also present at equilibrium.
At equilibrium, it was previously determined~[\onlinecite{aggregate}] that long nanoaggregates
contain more asphaltene molecules than resinous oil molecules and that the resin content is the smallest.
This explains why the anisotropy is larger for asphaltene molecules than for resinous oil molecules and
the smallest for resin molecules.
According to Fig.~\ref{fig:AggregateSizeVsShearRate}~(a), at temperature~$452$~K,
the nanoaggregates begin to break up at a shear rate very close to the critical shear rate
$\dot{\gamma}_c = 2\times 10^9$~s$^{-1}$. This indicates that the
increase of the molecular order parameter for shear rates higher than $\dot{\gamma}_c$
is due to the alignment of the single molecules in the flow direction. Again, this is consistent
with the shear thinning behavior observed at high shear rates.
Moreover, the existence of this intrinsic anisotropy in the sample could explain why the
value of the normal
stress ratio~$-\Psi_2/\Psi_1$ is different from what is found
for linear or branched polymer melts. Indeed, the normal
stress ratio~$-\Psi_2/\Psi_1$ is a measure of the anisotropy between
the $xx$- and $zz$- directions on the one hand and the $xx$-
and $yy$- directions on the other hand, which can be influenced
by an ordering of the system.

With the MD simulations which we carried out, we cannot know
whether the global anisotropy of the system is a finite size effect or if it
is also seen in macroscopic samples. In this last case,
bitumen could resemble the nematic phase of a discotic liquid crystal,
formed by the stacking of disk-shaped molecules~[\onlinecite{aguilera}].
Experimentally, it is known that there is a change of the supramolecular structure
of bitumen with the concentration of asphaltene molecules. At low concentrations,
small and isotropically distributed nanoaggregates can be seen. At larger concentrations,
the nanoaggregates gather in clusters. At even larger concentrations,
the clusters are all connected to form a network~[\onlinecite{mullins2011}].
This last stage could be representative of the simulations
carried out in this work.
The global anisotropy of the system was evaluated using
a molecular order tensor~$\textbf{Q}$ 
defined in a very similar way than the molecular
order tensor $S_m$, specified
for each molecule type in Eq.~\eqref{eq:Sm}.
The only differences are that the average is done over all aromatic molecules
instead of over one particular type
and that each molecule orientation is taken to be the unit vector normal
to the molecule plane instead of the principal direction determined
by the tensor of gyration.
The molecular order parameter $Q$ is then~$3/2$ of the largest eigenvalue of the
molecular order tensor~$\textbf{Q}$.
It is found to be equal to~$Q = 0.12\pm0.01$ at equilibrium~[\onlinecite{aggregate}].
This value of~$Q$ is sufficiently small, so that the shear viscosity can be treated as
as an isotropic scalar quantity.

To further investigate the effect of the anisotropy of the 
sample on the viscosity,
simulations were performed in the same shearing geometry as before and
with the same starting configurations except that the $y$- and $z$- coordinates are exchanged.
The simulations were carried out at a temperature of~$452$~K for the same shear rates and the same density.
The shear viscosity was analyzed and the following comparison was made.
Let $\eta_i$ and $\eta'_i$ denote the viscosities of the system starting with configuration~$i$
and the corresponding rotated configuration, respectively.
Let also $\bar{\eta}_i$ be the average $(\eta_i+\eta'_i)/2$.
The standard deviation~$s_{\text{rot}}$ due to the anisotropy of the configuration is quantified as:
\begin{equation}
s_{\text{rot}} = \sqrt{\frac{1}{N_{\text{conf}}} \sum_{i=1}^{N_{\text{conf}}} (\eta'_i - \bar{\eta}_i)^2},
\end{equation}
where $N_{\text{conf}} = 8$ is the total number of initial configurations considered.
A similar standard deviation~$s_{\text{conf}}$ comparing
different initial configurations and not an initial configuration with a rotation of itself
is defined as:
\begin{equation}
s_{\text{conf}} = \sqrt{\frac{1}{N_{\text{conf}}} \sum_{i\neq j} \bigl ( \eta_i - (\eta_i+\eta_j)/2\Bigr )^2},
\end{equation}
where $i$ and $j$ are two random different configurations,
chosen from the $8$ different configurations available. Exactly $N_{\text{conf}}$ pairs~$(i, j)$
are considered for the sake of symmetry with the definition of $s_{\text{rot}}$.
The variation of the ratio $r = s_{\text{rot}}/s_{\text{conf}}$
with the shear rate is displayed in Fig.~\ref{fig:molecularOrderParamVsShearRate}~(b).
The ratio of the two standard deviations is around~$1$ for all shear rates
indicating that the main source of difference between two configurations
is the anisotropy of the sample. The values of the viscosity given
in Fig.~\ref{fig:viscosityVsShearRate} should therefore be understood as an average
viscosity over different orientations. 

\begin{figure}
  \subfigure[]{\includegraphics[angle=-90, scale=0.25]{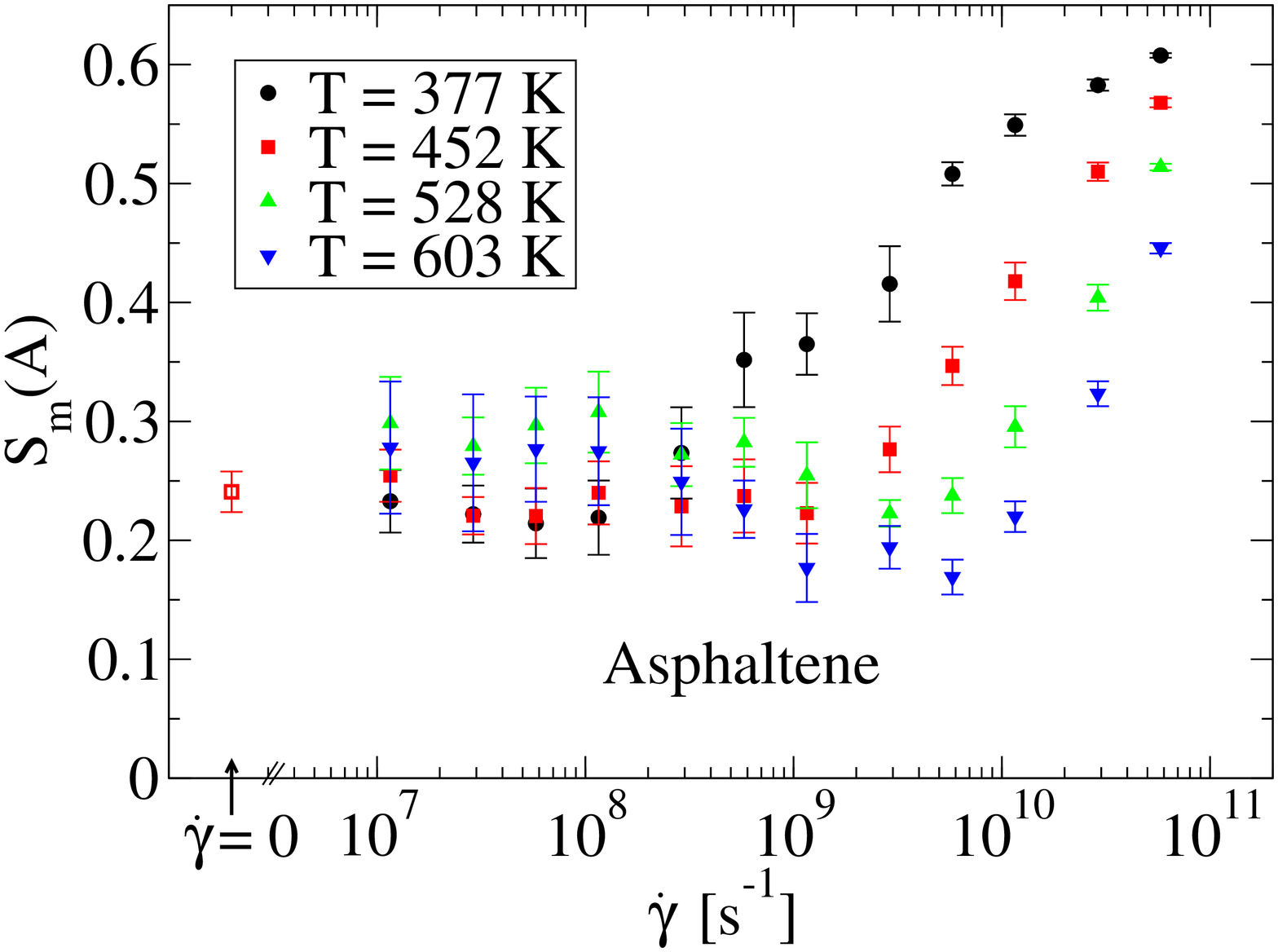}}
  \subfigure[]{\includegraphics[angle=-90, scale=0.25]{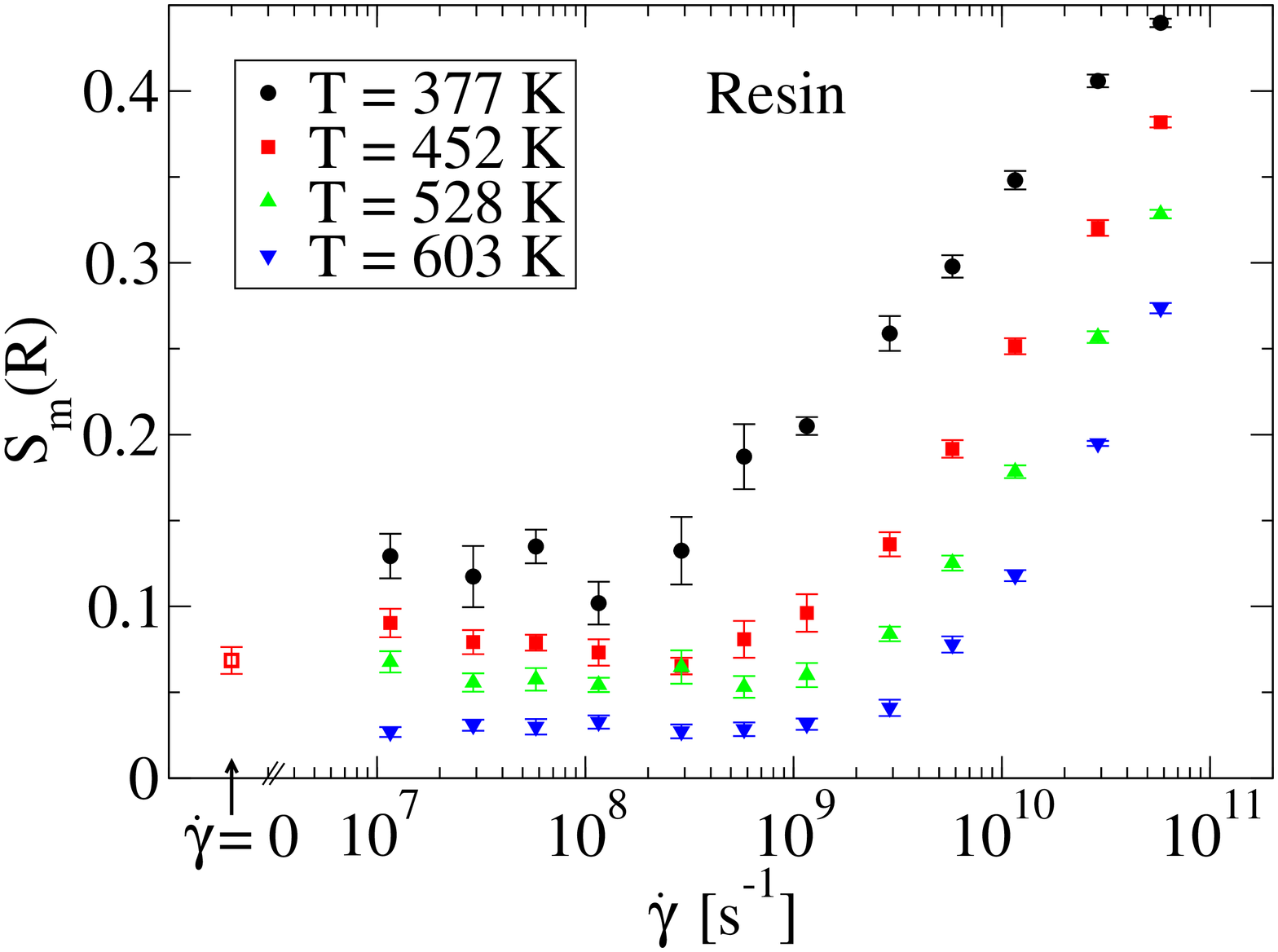}}
  \subfigure[]{\includegraphics[angle=-90, scale=0.25]{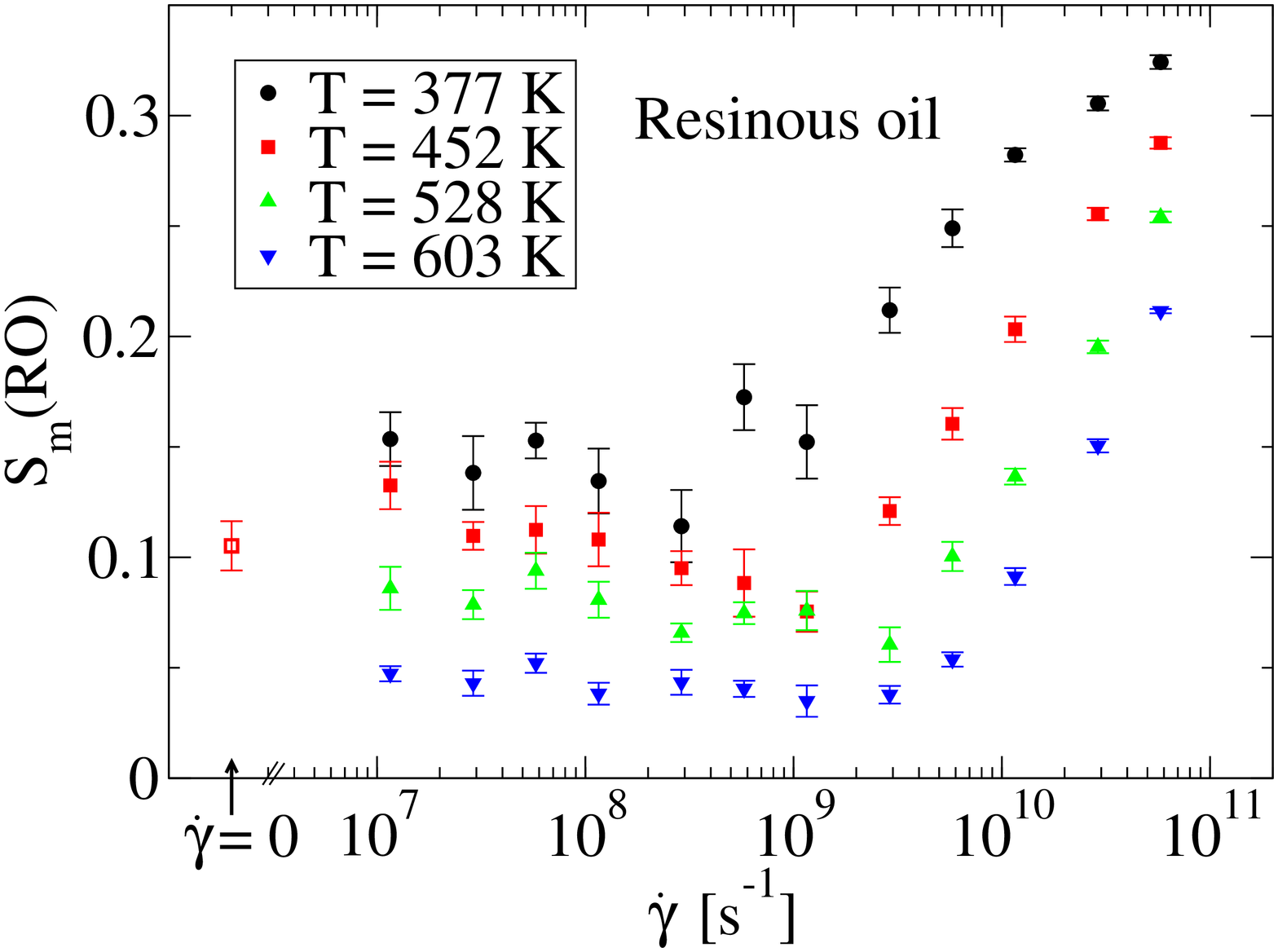}}
  \subfigure[]{\includegraphics[angle=-90, scale=0.25]{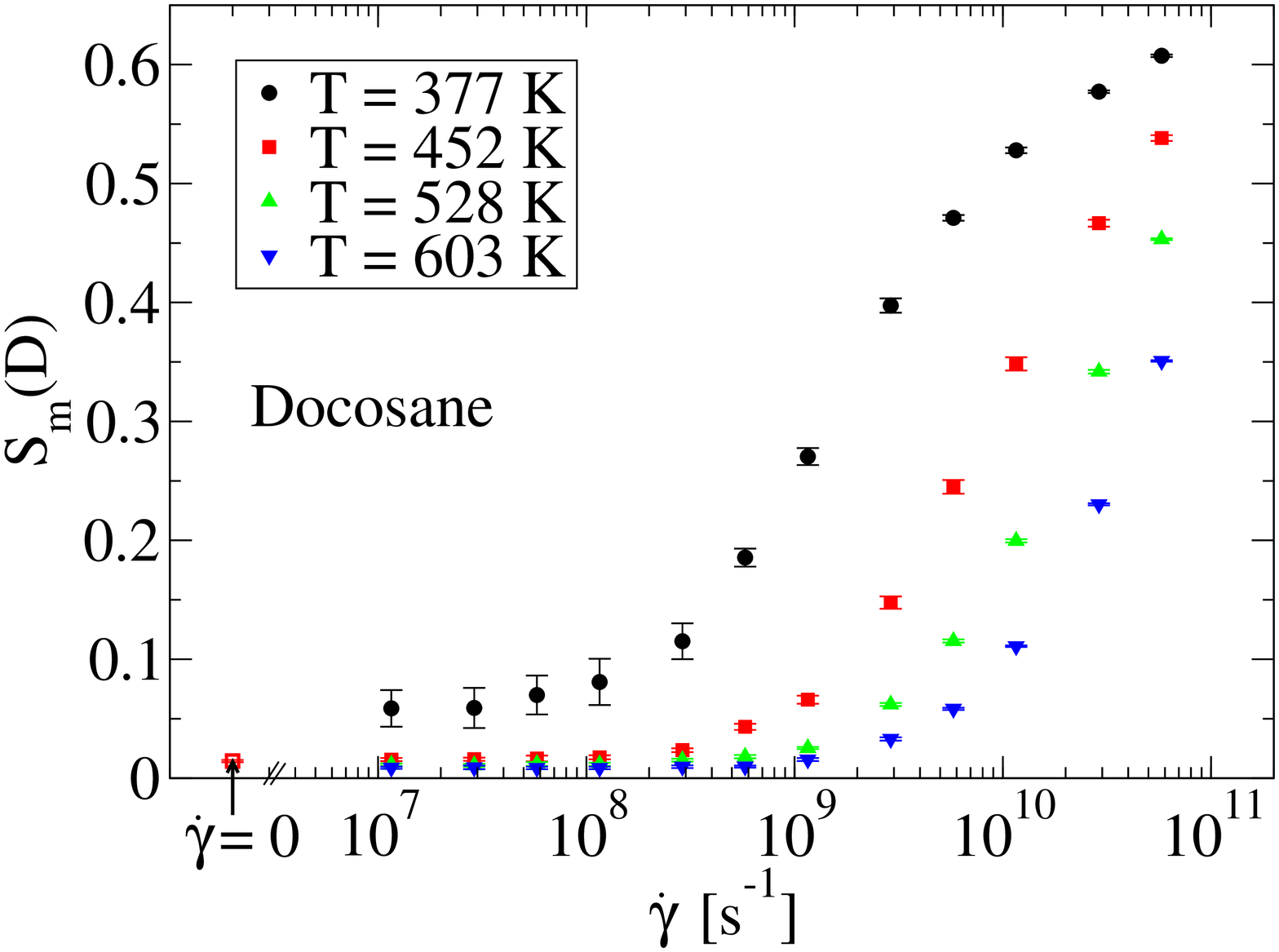}}

  \caption{\label{fig:molecularOrderParamDiffTemp}
(Color online)
Variation of the molecular order parameter~$S_m$ with shear rate~$\dot{\gamma}$
at different temperatures for asphaltene (a),
resin (b), resinous oil (c) and docosane (d) molecules.
}
\end{figure}

Figure~\ref{fig:molecularOrderParamDiffTemp} shows the variation of
the molecular order parameter of the four different molecule types
with shear rate for different temperatures.
From Fig.~\ref{fig:molecularOrderParamDiffTemp}, it is quite clear that
for all molecule types, the molecular order parameter
at a given high shear rate increases with decreasing temperature. This is due to the fact that
for a given shear rate at low temperatures the stress felt by the molecules is higher and
the alignment in the direction of the flow is more pronounced than at high temperatures.
At low shear rates, the behavior is quite different from one molecule type to another.
For docosane molecules, the molecular order parameter~$S_m(D)$ decreases with increasing
temperature, for the same reason as at high shear rates.
For resin and  resinous oil molecules the molecular order parameter also decreases
with increasing temperature at low shear rates. This is probably because
as the temperature increases, the resin and resinous oil molecules are more
often present as single molecules than in long nanoaggregates (see Fig~\ref{fig:AggregateSizeVsShearRate}~(d)),
which reduces their global ordering.
In contrast, asphaltene molecules stay in long nanoaggregates even at high temperatures and their molecular
order parameter is nearly independent of temperature at low shear rates.

An interesting feature can be noticed for the molecular order parameter~$S_m(A)$ of
asphaltene molecules at high temperatures. Looking closely at the data of Fig.~\ref{fig:molecularOrderParamDiffTemp}~(a)
at temperature~$603$~K, one can
note three different regimes with increasing shear rate: at low shear rates $S_m(A)$
is mostly constant, at intermediate shear rates $S_m(A)$ slightly decreases with
the shear rate, and finally at large shear rates it increases.
The intermediate shear rate range corresponds roughly to~$[3\times10^8, 6\times10^9]$~s$^{-1}$
for temperature~$603$~K. At these shear rates, the average number of molecules
in the nanoaggregates is still very close to its equilibrium value as can be seen in
Fig.~\ref{fig:AggregateSizeVsShearRate}~(a). Consequently, the decrease in the
molecular order parameter of the asphaltene molecules in this intermediate
shear rate range can be interpreted in the following way: the nanoaggregates
are still well formed, but the shear flow is strong enough to disturb the alignment of the
nanoaggregates with respect to each other causing the order parameter to locally drop.
At higher shear rates the nanoaggregates break, the molecules align in the shear direction
and the molecular order parameter increases again.
This intermediate drop of the molecular order parameter~$S_m(A)$ of asphaltene
molecules is also noticeable for temperature~$528$~K, but not for lower temperatures.
It might be because the nanoaggregates break before the shear flow is strong enough
to modify the alignment of the nanoaggregates with respect to each other.
The direct consequence of this reorganization of the nanoaggregates with respect to each other
on the behavior of the viscosity with the shear rate is unclear. 
In total, the fluid is shear-thinning in this range of shear rates, as
can be seen in Fig.~\ref{fig:viscosityVsShearRate}.
This global behavior could be mainly due to the alignment of the docosane
molecules in the shear direction, the order parameter of the resin and resinous oil
molecules being nearly constant in this range.

\subsubsection{Molecular alignment angle}
\label{sec:chim}
\begin{figure}
  \subfigure[]{\includegraphics[angle=-90, scale=0.25]{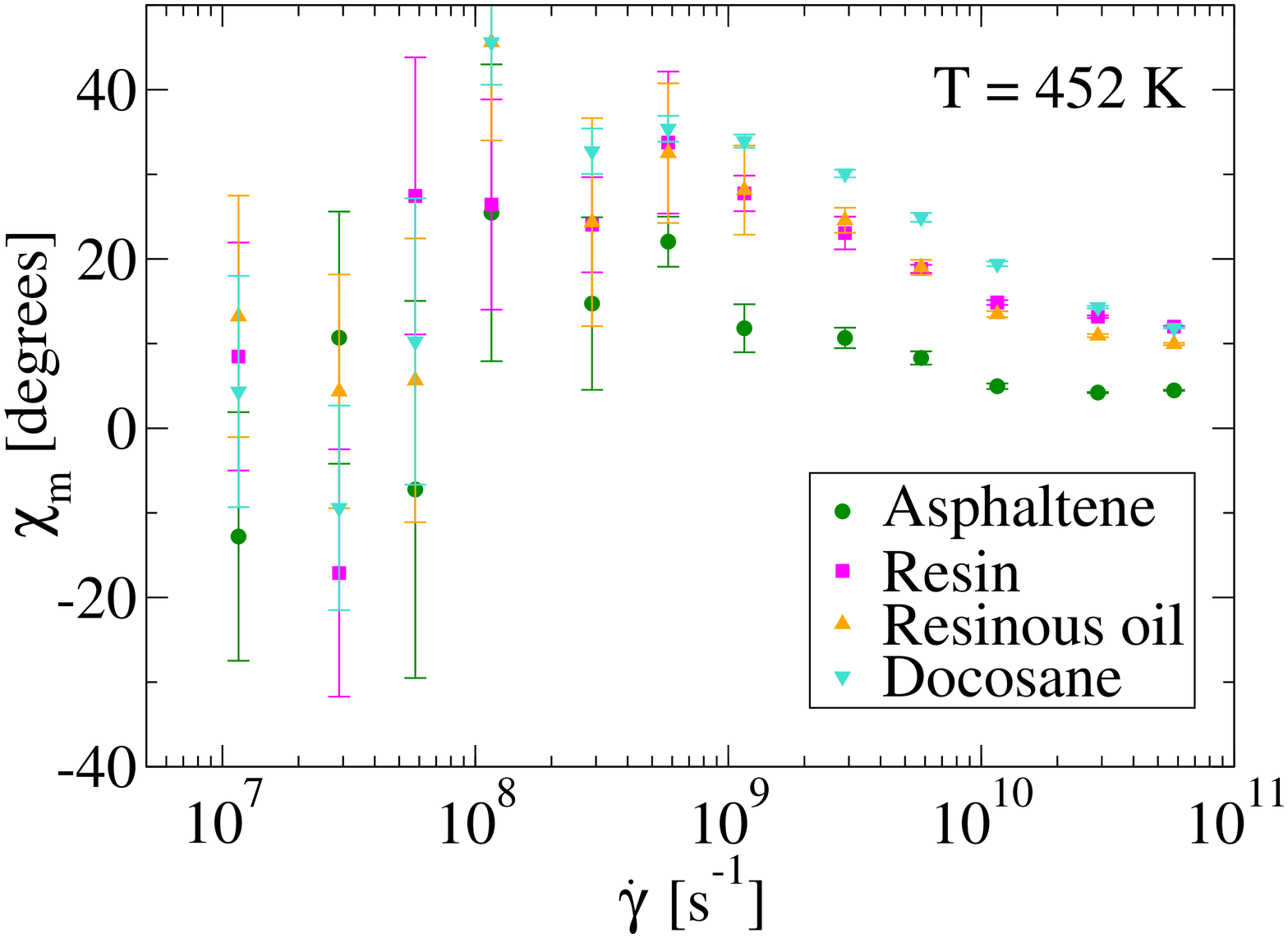}}\\
  \subfigure[]{\includegraphics[angle=-90, scale=0.25]{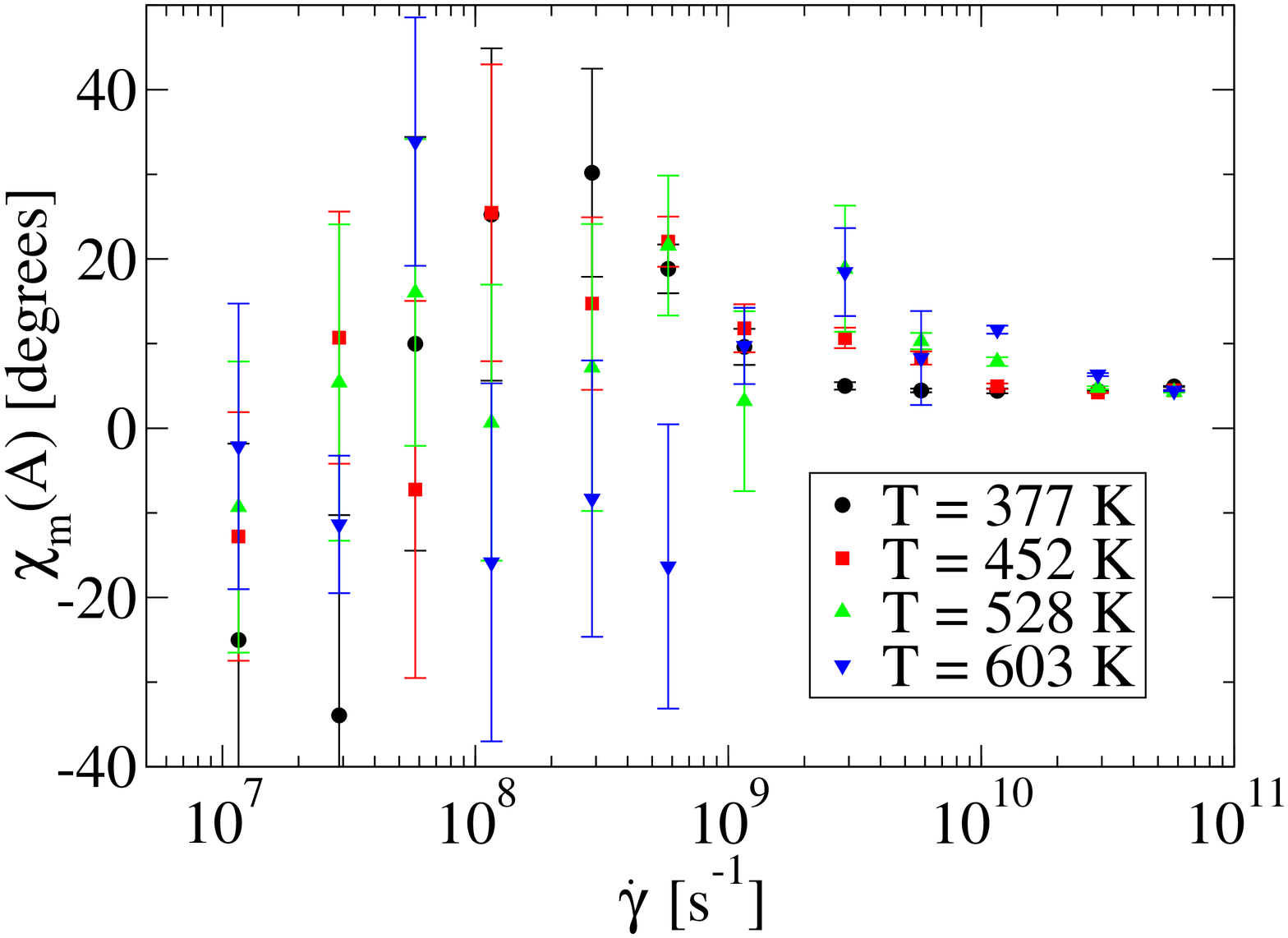}}
  \subfigure[]{\includegraphics[angle=-90, scale=0.25]{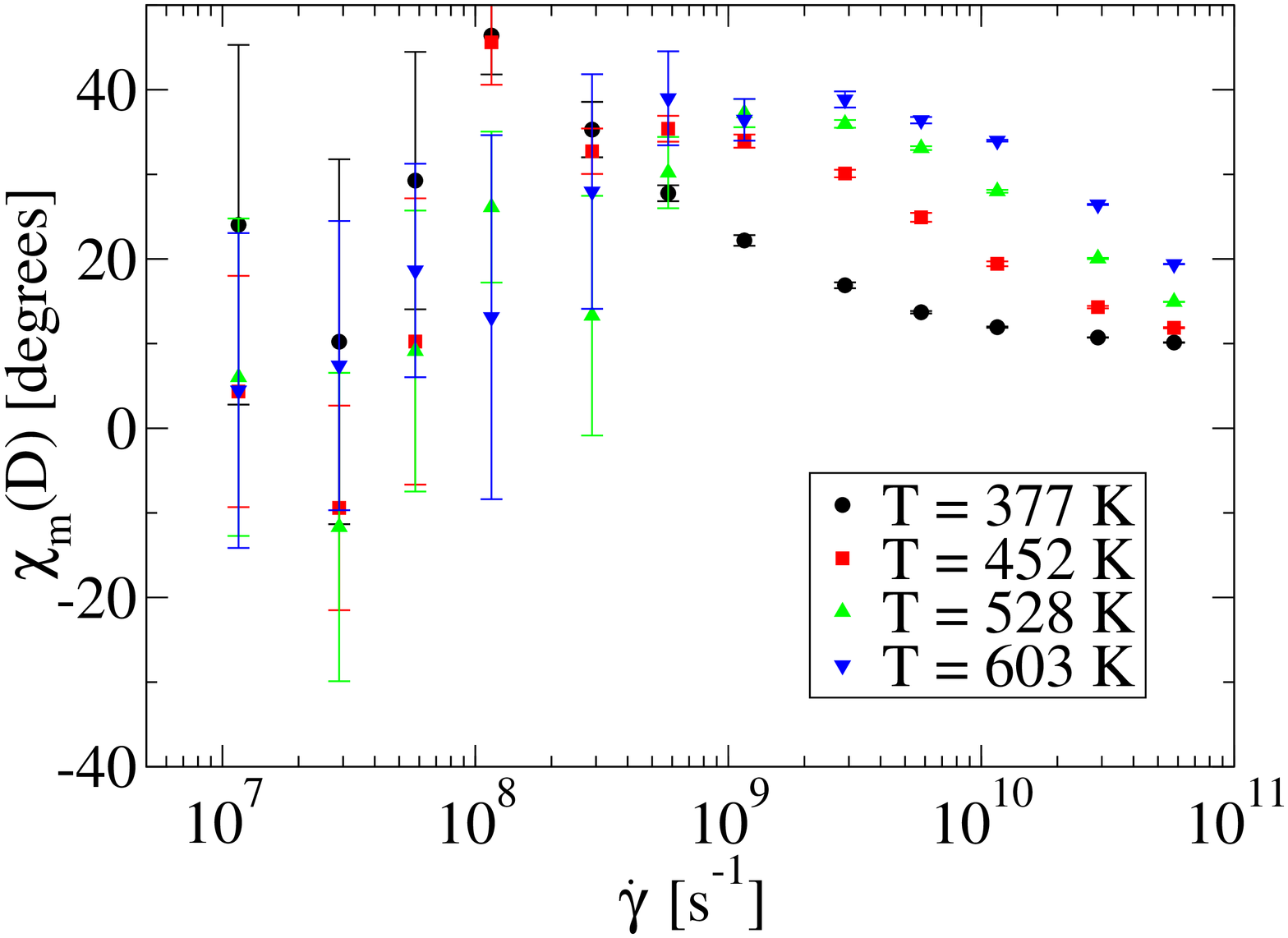}}
  \caption{\label{fig:molecularAngleVsShearRate}
(Color online)
Variation of the molecular alignment angle~$\chi_m$ with shear rate~$\dot{\gamma}$
for three cases: at temperature~$452$~K and
for different molecule types (a), for asphaltene molecules at different temperatures (b), and
for docosane molecules at different temperatures (c).
}
\end{figure}

The variation of the molecular alignment angle~$\chi_m$ is plotted
for each molecule type versus
the shear rate~$\dot{\gamma}$ at temperature~$452$~K in Fig.~\ref{fig:molecularAngleVsShearRate}~(a).
The typical behavior for a polymer solution under shear is the following:
in the Newtonian regime, the molecular alignment angle is around~$45$~degrees and
in the non-Newtonian regime, the molecular alignment angle decreases until reaching zero.
The value of~$45$~degrees is expected in the Newtonian regime, due to
the nonuniform distribution of angular velocity of the molecules in the shear flow~[\onlinecite{le}].
At larger shear rates, the molecules align to the shear flow resulting in a vanishing
molecular alignment angle.
This general behavior is roughly seen for all molecule types in Fig.~\ref{fig:molecularAngleVsShearRate}~(a),
though the error bars are very large at low shear rates.
The large error bars at low shear rates can be
explained by the fact that at these shear rates the nanoaggregates are well formed and
tend to align with respect to each other,
giving rise to the slight anisotropy of the whole sample.
The alignment of the nanoaggregates with respect to each other causes the set of the angles
with the shear direction to
be ill-spanned compared to an isotropic material.
The error bars decrease when the nanoaggregates are ruptured. This happens at quite large
shear rates, when the single molecules begin to align in the flow direction, so that
the angle of~$45$~degrees is not clearly visible.
Furthermore, at high shear rates, the values of the molecular alignment angle~$\chi_m$
for different molecule types
are ordered according to the molecular weight, the molecules with the highest
weight having the lowest alignment angle. It is expected as 
the difference in velocity from one end of the molecule to the other
is larger for large molecules than for small molecules, giving rise to an
alignment in the flow direction at lower shear rates.

Finally, Figs.~\ref{fig:molecularAngleVsShearRate}~(b) and (c)
show the molecular alignment angle of asphaltene and docosane molecules, respectively,
at different temperatures. 
The molecular alignment angle of resin and resinous oil molecules  at different temperatures is not
displayed for the sake of readability and resembles that of the asphaltene molecules
at different temperatures.
For both asphaltene and docosane molecules, the lower the temperature,
the closer to zero the angle to the flow direction for a given high shear rate.
This is consistent with the molecular order parameter being higher for low temperatures
than for high temperatures at a given high shear rate.
The error bars on the alignment angle of asphaltene molecules are large up to
very high shear rates for high temperatures, consistent with the nanoaggregates
being maintained up to these high shear rates.
Due to the large error bars, the existence of the intermediate range of shear rates
identified in Fig.~\ref{fig:molecularOrderParamVsShearRate}~(a) and corresponding to
the shear flow disturbing the alignment of the nanoaggregates with respect to each other,
does not have a clear effect on the alignment angle of the asphaltene molecules in the flow direction.

\subsection{Intramolecular structure}
\label{sec:intramol}

\begin{figure}
   \subfigure[]{\includegraphics[angle=-90, scale=0.25]{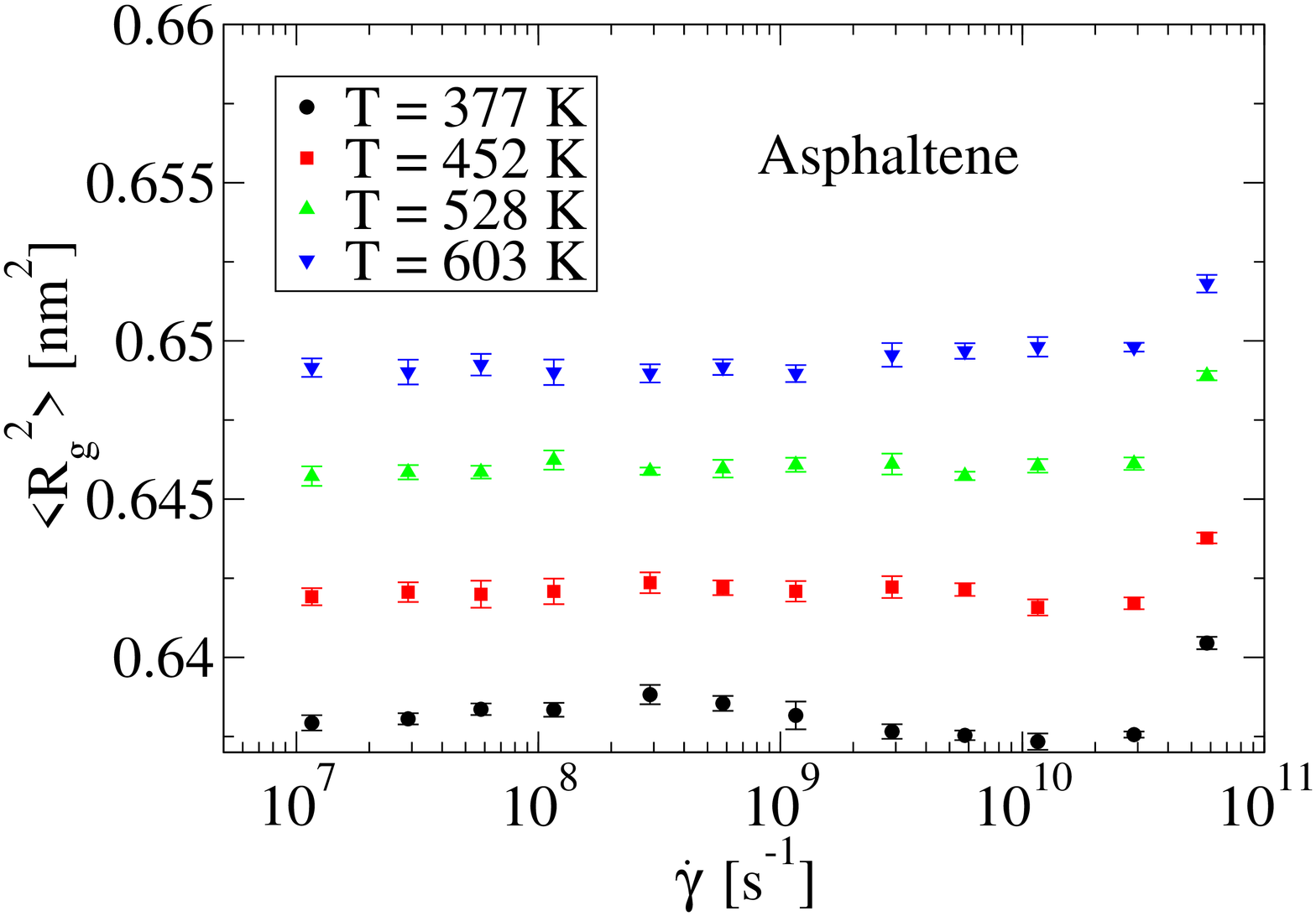}}
   \subfigure[]{\includegraphics[angle=-90, scale=0.25]{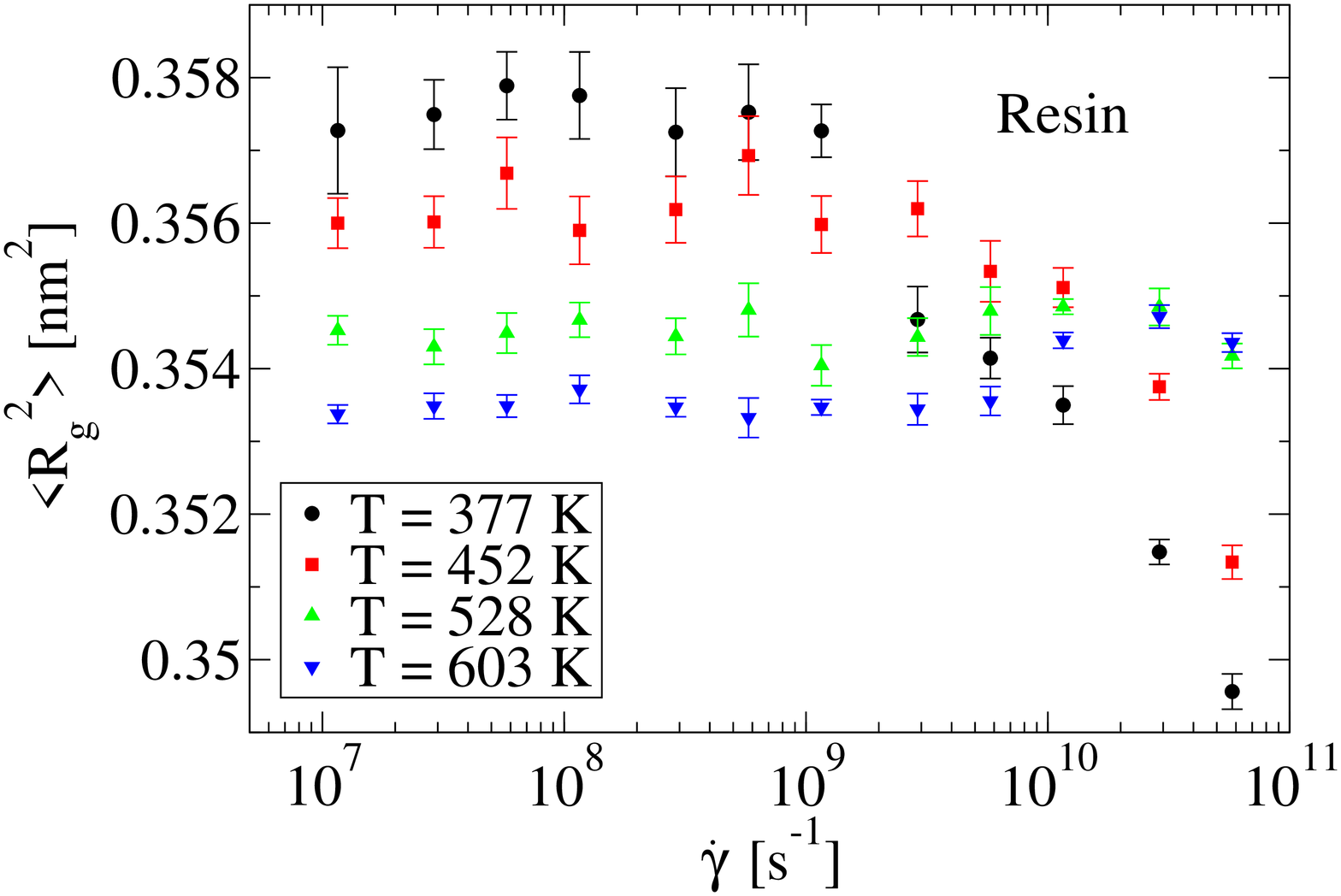}} \\
   \subfigure[]{\includegraphics[angle=-90, scale=0.25]{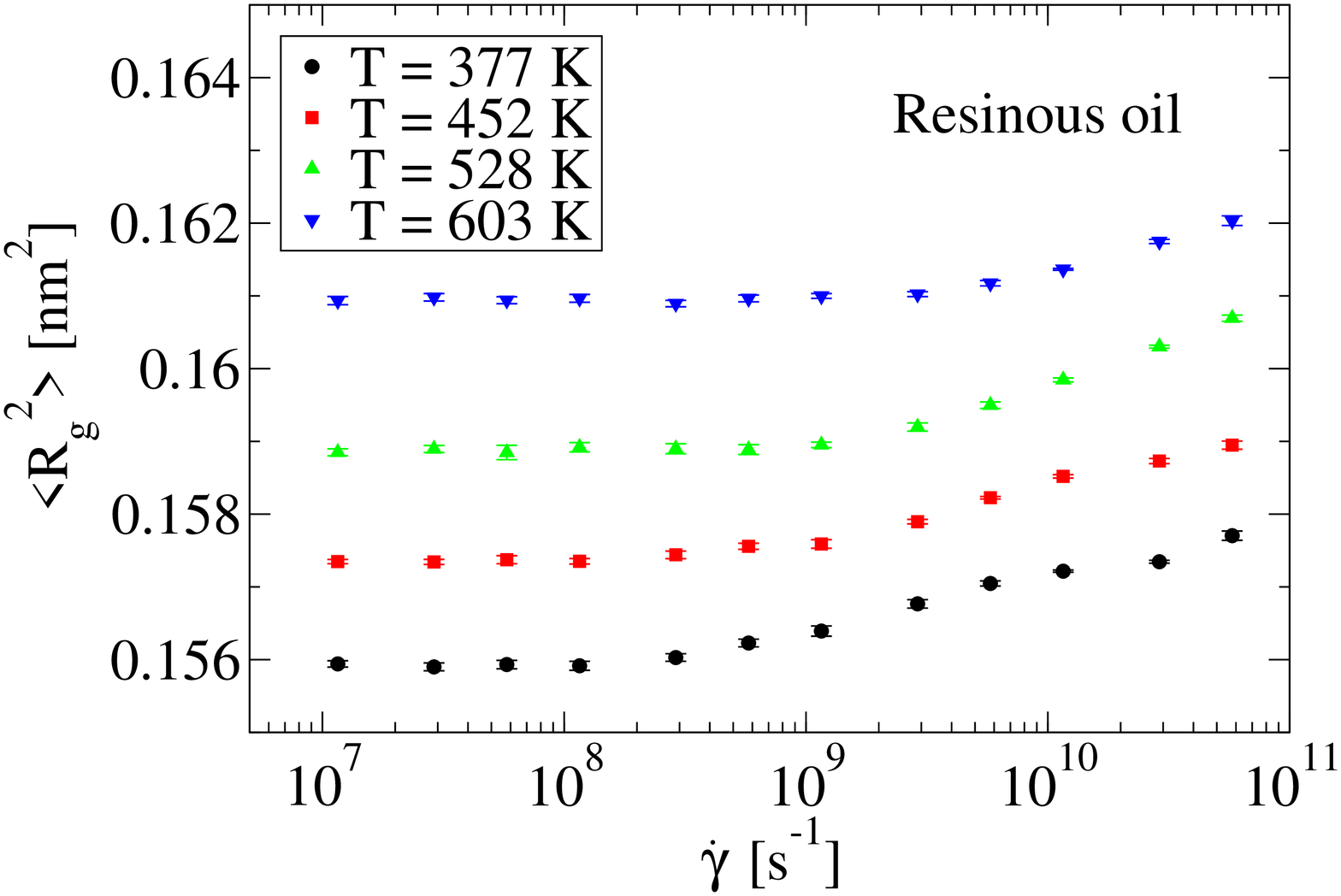}}
   \subfigure[]{\includegraphics[angle=-90, scale=0.25]{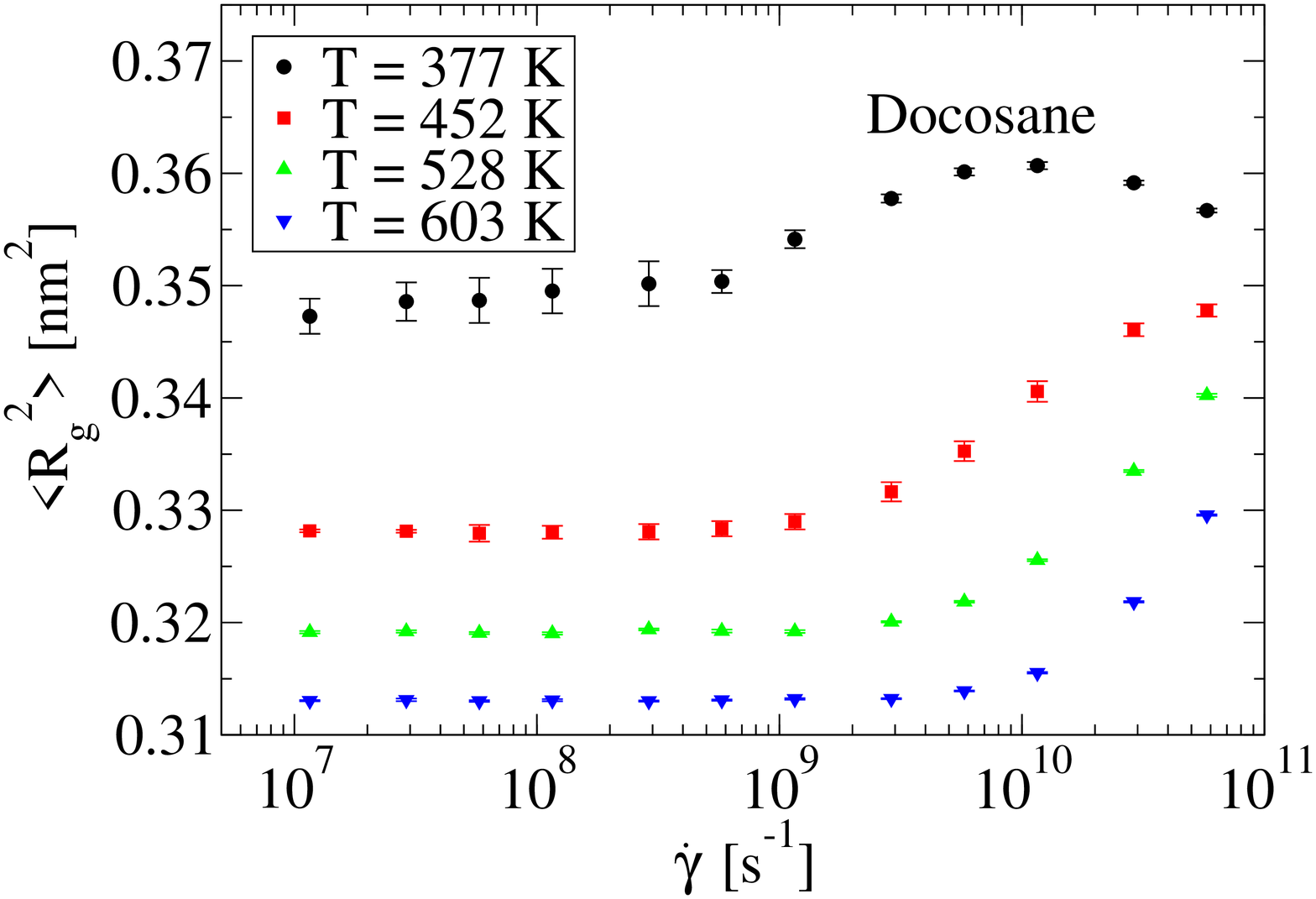}}
  \caption{\label{fig:radiusOfGyrationVsShearRate}
(Color online) 
Variation of the radius of gyration with the shear rate for asphaltene (a), resin (b), resinous
oil (c) and docosane (d) molecules at different temperatures.
}
\end{figure}

The intramolecular structure can be characterized by the radius of gyration of each molecule
type.
The radius of gyration is defined here as $\langle R_g^2\rangle = \text{trace}\langle\mathbf{R}^2_g\rangle$,
where $\mathbf{R}^2_g$ is the tensor of gyration given in Eq.~\eqref{eq:tensorOfGyration} and
$\langle\cdot \rangle$ is an average over molecules of the same type and over time.
Figure~\ref{fig:radiusOfGyrationVsShearRate} shows the variation
of the radius of gyration of each molecule type with the shear rate for different temperatures.
The variation of the radius of gyration with temperature and with shear rate is very different from
one molecule type to another. 
For asphaltene and resinous oil molecules, Figs.~\ref{fig:radiusOfGyrationVsShearRate}~(a)
and~(c) respectively, the radius of gyration barely changes with the shear rate.
It is expected as nearly all carbons in these molecules are aromatic creating
a stiff intramolecular structure.
However, at the largest shear rates the radius of gyration increases 
slightly indicating a stretching of
the bonds.
For both molecule types the radius of gyration increases with increasing
temperature, indicating a stretch of the stiff bonds due to a higher velocity
of each bead.

For docosane and resin molecules, Figs.~\ref{fig:radiusOfGyrationVsShearRate}~(b)
and~(d) respectively, the trends are very different.
Both molecule types can possess a maximum of their radius of gyration at a given
shear rate. For linear molecules such as docosane molecules, this behavior
is known~[\onlinecite{khare, bosko2004}] and is attributed to a stretching of the bonds in the flow direction
for intermediate shear rates and a coiling up of the whole molecule
at very high shear rates when each molecule tumbles more often.
The maximum in the radius of gyration of linear molecules is usually associated with
a clear minimum of the pressure of the system~[\onlinecite{khare, bosko2004}].
It is not seen in the case of the Cooee bitumen model, probably because
other molecules play a part in the value of the pressure as mentioned in Sec.~\ref{sec:normalPressure}.
We attribute the maximum seen for resin molecules at a given shear rate to a similar effect,
because
resin molecules have some long linear alkyl chains aside from the
aromatic plane, in contrast to resinous oil and asphaltene molecules.
Finally, for both docosane and resin molecules, the radius of gyration
decreases with increasing temperature. This could also be an effect of the
larger velocity of each bead, causing coiling of the molecules when the
bonds are less stiff than in pure aromatic molecules.

\begin{figure}
  \subfigure[]{\includegraphics[angle=-90, scale=0.25]{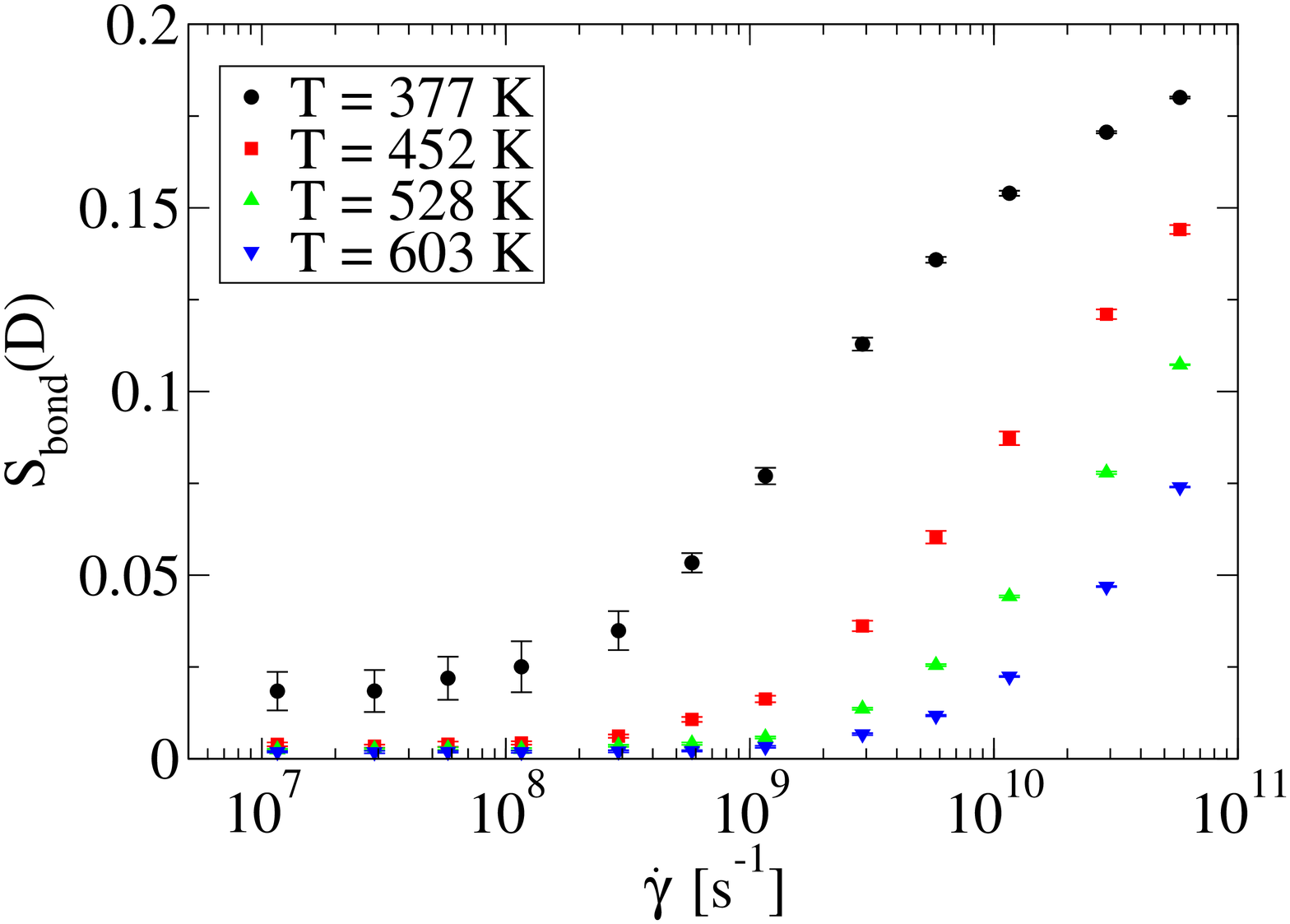}}
  \subfigure[]{\includegraphics[angle=-90, scale=0.25]{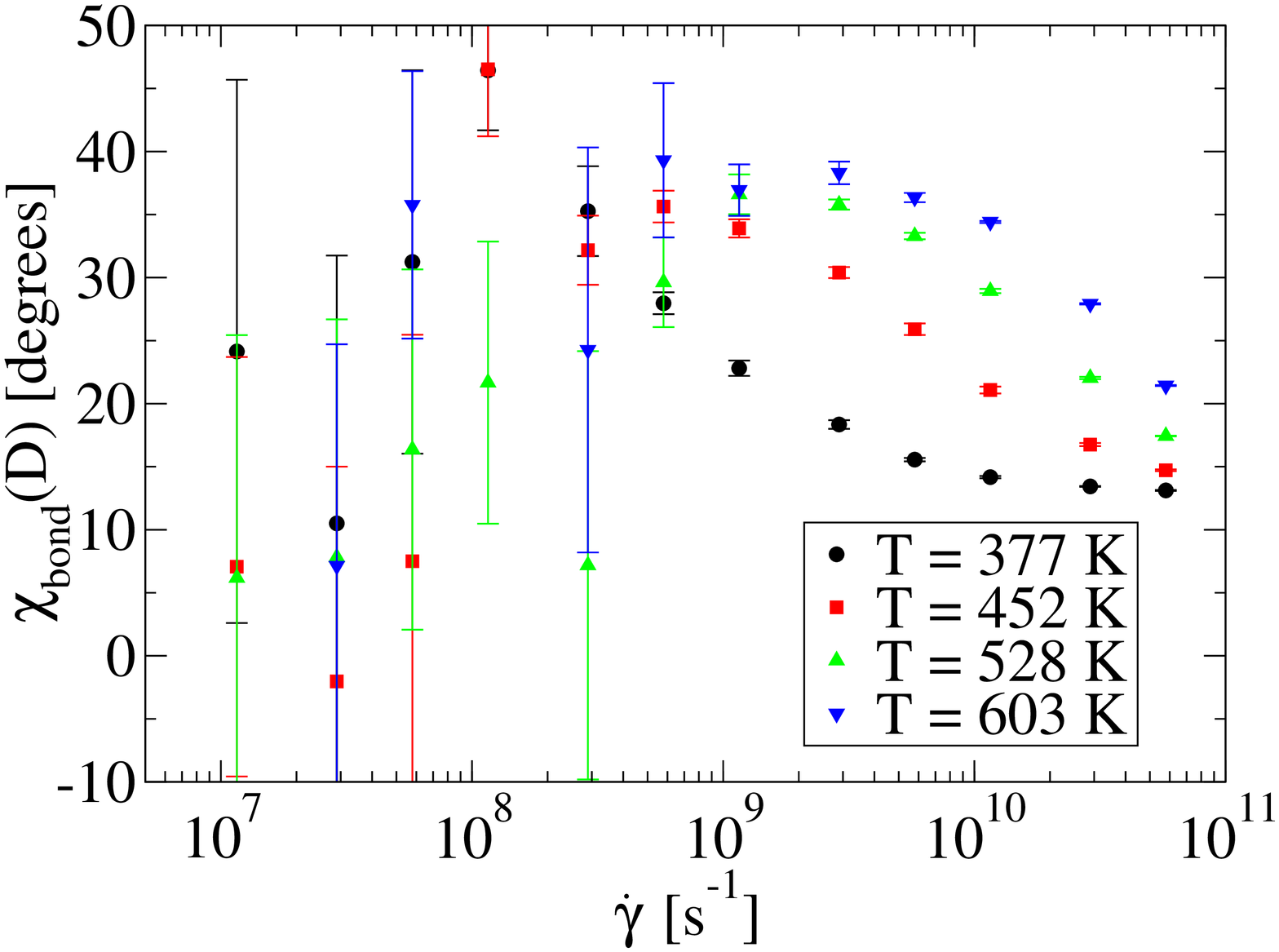}}
  \caption{\label{fig:bondOrderParamVsShearRate}
(Color online)
Variation of the bond order parameter~$S_{\text{bond}}$ (a) 
and the bond alignment angle~$\chi_{\text{bond}}$ (b) with shear rate~$\dot{\gamma}$
for docosane molecules at different temperatures.
}
\end{figure}

Another informative quantity which can be defined to describe the intramolecular structure
is the bond order tensor~$S_{\text{bond}}$. It is defined for docosane molecules as:
\begin{equation}
\mathbf{S}_{\text{bond}} = \frac{1}{N_b} \sum_{\alpha=1}^{N_b-1} \Bigl \langle \mathbf{v}_{\alpha}\otimes\mathbf{v}_{\alpha} -\frac{1}{3}\mathbf{I}\Bigr \rangle,
\end{equation}
where $N_b$ is the number of bonds in a docosane molecule,
$\langle\cdot\rangle$ is an average over docosane molecules and over time, and $\mathbf{v}_{\alpha}$
is the unit vector between neighboring atoms~$\alpha$ and~$\alpha+1$ given as:
\begin{equation}
\mathbf{v}_{\alpha} = \frac{\mathbf{r}_{\alpha+1}-\mathbf{r}_{\alpha}}{|\mathbf{r}_{\alpha+1}-\mathbf{r}_{\alpha}|}.
\end{equation}
In a similar way as for the molecular order tensor~$\mathbf{S}_m$,
a bond order parameter~$S_{\text{bond}}$ and a bond alignment angle~$\chi_{\text{bond}}$
can be defined from the bond order tensor~$\mathbf{S}_{\text{bond}}$.
These two quantities measure the extent of the alignment of intramolecular
bonds to the shear direction and the angle between them.
Figures~\ref{fig:bondOrderParamVsShearRate}~(a) and (b)
show the variation of the bond order parameter~$S_{\text{bond}}$ and the bond
alignment angle~$\chi_{\text{bond}}$ of docosane molecules with the
shear rate for different temperatures.
Both quantities have similar variations with shear rate and temperature
as their molecular counterparts. As the shear rate increases,
the bonds inside the docosane molecules become more and more aligned in the direction of the shear.
This effect gets stronger as the temperature decreases. It is also
consistent with the shear-thinning behavior observed in Sec.~\ref{sec:viscosity}.

The bond order tensor associated with aromatic molecules
is less well defined, because the bonds form rings and are not clearly oriented
from one end of the molecule to the other. 
That is why it is not calculated here.
However, as shown
by the variation of the radius of gyration of the aromatic molecules
with the shear rate, the stretching of the molecules 
is less pronounced for stiff aromatic molecules than for
linear ones. Changes in the intramolecular structure of the
aromatic molecules probably contribute very little to
the shear-thinning behavior generally observed.

\section{Summary and discussion}
\label{sec:discussion}

Non-equilibrium molecular dynamics
simulations under shear have been performed for the first time on a bitumen mixture.
They reveal both the non-Newtonian regime of such mixtures and
their microscopic structure.
The non-Newtonian regime is characterized by a shear rate dependent
viscosity, decreasing normal stress coefficients, and a monotonically increasing pressure.
The Cooee bitumen is found to be shear-thinning at all temperatures
investigated in line with most
experimental results on bitumen~[\onlinecite{sybilski}].
The shear-thinning behavior is in agreement with the variation of some
characteristics of the inter- and
intramolecular structure with the shear rate:
\begin{itemize}
\item The saturated hydrocarbon molecules,
which can be seen as solvent to the nanoaggregates, align with respect to each other and align their bonds
in the shear direction, at all shear rates.
\item At high shear rates, the nanoaggregates disintegrate and the single aromatic
molecules align in the shear direction.
\end{itemize}
Some characteristics of the inter- and 
intramolecular structure of the Cooee bitumen do not change with the shear rate
and do not seem to prevent the shear-thinning behavior from being established.
They are:
\begin{itemize}
\item Until a critical shear rate, dependent on the temperature, the nanoaggregate
size distribution is unchanged with increasing shear rate.
\item The intramolecular structure of the aromatic molecules, composing the
nanoaggregates, is mainly constant up to quite high shear rates.
\end{itemize}
Finally at high temperatures, there is a domain of intermediate shear rates,
where the nanoaggregates still have their equilibrium size,
but where the molecular order parameter of asphaltene molecules decreases
with increasing shear rate. The decrease of the molecular order parameter
indicates a perturbation of the alignment of the nanoaggregates with
respect to each other, due to steric hindrance, by the shear flow.
The effect of this local reorganization on the
variation of the viscosity with the shear rate is unclear and could be
masked by the inter- and intramolecular alignment of the solvent molecules.

In addition, the variation of the rheological properties and of the
corresponding molecular structure with temperature was studied.
As expected, the viscosity decreases with increasing temperature, following
an Arrhenius law. 
The nanoaggregate size decreases with increasing temperature, which
also contributes to the decrease of viscosity with temperature.
The increase in viscosity at low temperatures and at a given shear rate
explains some
of the characteristics of the molecular structure:
\begin{itemize}
\item The solvent molecules begin to align
in the shear direction at lower shear rates at low temperatures than at high temperatures.
\item The nanoaggregates break up at a lower shear rate at low temperatures than at high
temperatures. This is also due to the nanoaggregates being longer
at low temperatures than at high temperatures.
\end{itemize}

The study of the structure and rheology of the bitumen mixtures at different temperatures
enables us to compare the behavior
of a bitumen mixture with other
complex systems such as colloidal suspensions, polymer solutions and associative polymer networks.
The comparison with these complex systems is relevant because
they have similarities with a bitumen mixture. 
Indeed, the Cooee bitumen
mixture can be seen as a solution of partly flexible nanoaggregates
of various sizes~[\onlinecite{aggregate}], with a branched structure~[\onlinecite{aging}],
in a saturated hydrocarbon solvent, as was already mentioned in Sec.~\ref{sec:aggregateSize}.
Furthermore, 
the nanoaggregate size depends on temperature and shear rate
as was seen in Sec.~\ref{sec:molStruc}. 

It is known that both the linear rheological properties such as the
zero shear rate viscosity and the non-linear rheological properties
of colloidal suspensions and polymer solutions
depend on the size, the shape and the polydispersity of
the dispersed objects~[\onlinecite{genovese, mansfield2013, winkler}].
In the limit of a dilute solution of such objects,
a lot is known on the relation between the intrinsic viscosity~$[\eta]$,
defined by $\eta_0 = \eta_{0s} + [\eta] \phi +o(\phi)$, where $\eta_{0s}$
is the zero shear rate viscosity of the solvent and $\phi$ the volume fraction
of objects, and the shape of the colloids or the topology of the
polymers~[\onlinecite{mansfield2013}].
The shear-thinning
behavior at high shear rates can also be related to the topology
of the polymers, though for complicated topology such as
dendrimers, quantitative models are not available yet to predict
the exponent of the decay of the viscosity in the limit of infinite shear rate~[\onlinecite{winkler}].
At high volume fractions of objects, a higher order expansion of these linear and non-linear
rheological properties with the volume fraction is needed,
because the interactions between the colloids
or polymers cannot be neglected~[\onlinecite{coussot}].
Bitumen mixtures typically fall in this range,
which makes a direct and quantitative comparison with what is known of the
rheological properties of dilute and semi-dilute polymer solutions of different topologies difficult.

Another characteristic of bitumen which makes a direct comparison with
colloidal suspensions or polymer solutions difficult is the fact that
nanoaggregates are maintained by weak
interactions and their size depends on temperature and shear rate.
This is usually not the case for colloids
and for polymer solutions.
Theoretical~[\onlinecite{vaccaro, indei, tripathi, koga}] and
numerical~[\onlinecite{kroger96, milchev, bedrov, padding, huang, li}]
studies have been carried out on the linear and non-linear rheological properties
of solutions of associative polymers, where the size of the assemblies also depends
on temperature and shear rate. There is for example
a growing literature on the rheology of 
telechelic polymer networks~[\onlinecite{vaccaro, bedrov, indei, tripathi, koga}]
and of wormlike micelles~[\onlinecite{kroger96, milchev, padding, huang}].
However, the knowledge in this field is still scarce and very specific to the precise
topology of self-assembly studied~[\onlinecite{li}]. It is consequently not so
easily transferable to bitumen.

Nevertheless, the current knowledge on the rheology of colloidal suspensions, polymer
solutions and associative polymer solutions defines a framework which can be used to better
understand bitumen. The size, the polydispersity and the topology of the nanoaggregates
as well as the interaction between the nanoaggregates are expected to be the important parameters
controlling the rheological properties of bitumen. We hope that the numerical
study carried out in this paper can be used as a step towards an analytical
modeling of the rheological properties of bitumen based on these parameters.

\section*{Acknowledgements}
This work is sponsored by the grants 1337-00073B and 1335-00762B
of the Danish Council for Independent Research $|$ Technology and Production
Science. It is in continuation of the Cooee project (CO$_2$ emission reduction
by exploitation of rolling resistance modeling of pavements), sponsored by
the Danish Council for Strategic Research. The centre for viscous liquids dynamics
"Glass and Time" is supported by the Danish National Research Foundation's grant DNRF61.

\appendix

\section{\label{operatorsplitting}Operating splitting algorithm for molecular SLLOD}

We have implemented a molecular version of the operator-splitting algorithm described in Ref~\onlinecite{pan}. 
To our knowledge, it is the first time that this algorithm is adapted
to molecular systems. The implementation involves a two-step process: (1) the equations of motion of the molecular centers of mass are solved; (2) the (relatively simple) equations of motion for the particles' positions and velocities relative to the molecular centers of mass are solved and combined with the solutions for centers of mass in order to update the atomic coordinates and velocities. Note that here velocities are the so-called peculiar velocities, i.e. velocities relative to the streaming velocity.

The operator splitting method is an approximation based on a factorization of time propagator $U$ generated by the system's Liouville operator $iL$. The factorization is denoted $\mathbf{A} \mathbf{B}_1 \mathbf{B}_2 \mathbf{B}_1 \mathbf{A} $ where $\mathbf{A}$ is the operator associated with evolution of the coordinates at fixed momenta, $\mathbf{B}_1$ is that associated with evolution of the momenta due to the flow only, and $\mathbf{B}_2$ is that associated with the evolution of momenta in the absence of flow, at fixed particle coordinates. Exact solutions of each set of equations of motion are given by Pan {\it et al}\cite{pan}. Note that $\mathbf{A}$ and $\mathbf{B}_1$ appear twice and thus should be integrated over a half time-step each time, although the first instance of $\mathbf{A}$ can be ``wrapped'' around and put at the end, at the expense of having the positions and velocities and positions a half-step out of sync. The isokinetic thermostat is included in each factor, so that the kinetic energy is exactly conserved (up to round-off error).

For the molecular case the same factorization is made and the above two step procedure is applied for each factor in the approximate propagator. The equations of motion for operator $\mathbf{A}$ are, for our choice of simple shear deformation (where $i$ is a molecular index, $\alpha$ is a particle index within a molecule, and quantities with only the former are center-of-mass quantities):
\begin{align}
\mathbf{\dot{r}}_{i \alpha} &=  \frac{\mathbf{p}_{i \alpha}}{m_{i \alpha}} + \dot\gamma y_i \hat x = \mathbf{v}_{i \alpha} + \dot\gamma y_i \hat x, \label{eq:atomic_EOM_A}\\
\mathbf{\dot{p}}_{i \alpha} &= 0
\end{align}
For later convenience we introduce the atomic and molecular velocities~$ \mathbf{v}_{i \alpha}$ and $ \mathbf{v}_{i}$ respectively. Focusing on the positions, we multiply by the mass fraction~$m_{i \alpha}/M_i$ of the particle $\alpha$ in molecule~$i$ and sum over $\alpha$ to get the equation of motion for the center of mass position:
\begin{equation}
\mathbf{\dot{r}}_{i} =  \frac{\mathbf{p}_{i}}{M_i} + \dot\gamma y_i \hat x =  \mathbf{v}_{i} + \dot\gamma y_i \hat x. \label{eq:CM_EOM_A}
\end{equation}
This is identical to the position part of the $\mathbf{A}$ operator in Ref.~\onlinecite{pan} (re-interpreting the index $i$ in their equations as a molecular index), and has the same solution for a time increment $\Delta t$ (although incorrectly given in that paper; the correct expressions can be found in Ref.~\onlinecite{Zhang/others:1999}):

\begin{align}
x_i(t+\Delta t) &= x_i(t) + \Delta t(p_{xi}(t)/M_i + \dot\gamma y_i(t)) + (1/2M_i)\Delta t^2 \dot\gamma p_{yi}(t) \label{eq:sol_A_CM_x} \\
y_i(t+\Delta t) &= y_i(t) + \Delta t(p_{yi}(t)/M_i)  \\
z_i(t+\Delta t) &= z_i(t) + \Delta t(p_{zi}(t)/M_i)\label{eq:sol_A_CM_z}
\end{align}
Knowing the evolution of the center of mass position during the time step $\Delta t$, we can now consider the equations of motion for the relative coordinates of the particles within the molecule by subtracting Eq.~\eqref{eq:CM_EOM_A} from Eq.~\eqref{eq:atomic_EOM_A}, giving
\begin{equation}
\frac{d}{dt} \left( \mathbf{r}_{i\alpha} - \mathbf{r}_i \right)
= \mathbf{p}_{i\alpha}/m_{i\alpha} - \mathbf{p_i}/M_i = \mathbf{v}_{i\alpha} - \mathbf{v}_{i},
\end{equation}
where the right side is the particle's velocity relative to the molecule's center of mass and is constant for the $\mathbf{A}$ equations of motion. The solution is therefore (using velocities instead of momenta)
\begin{equation}
 \mathbf{r}_{i\alpha}(t+\Delta t) - \mathbf{r}_{i}(t+\Delta t) =  \mathbf{r}_{i\alpha}(t)  -  \mathbf{r}_{i}(t) + \Delta t( \mathbf{v}_{i\alpha}(t) - \mathbf{v_i}(t)),
\end{equation}
which, upon substituting Eqs.~\eqref{eq:sol_A_CM_x}-\ref{eq:sol_A_CM_z}, gives
\begin{align}
x_{i\alpha}(t+\Delta t) &= x_{i\alpha}(t) + \Delta t(v_{xi\alpha}(t) + \dot\gamma y_i(t)) +  1/2 \Delta t^2 \dot\gamma v_{yi}(t) \\
y_{i\alpha}(t+\Delta t) &= y_{i\alpha}(t) + \Delta t(v_{yi\alpha}(t)) \\
z_{i\alpha}(t+\Delta t) &= z_{i\alpha}(t) + \Delta t(v_{zi\alpha}(t))
\end{align}

The same procedure is applied to the equations of motion for the $\mathbf{B}_1$ operator, which for the atomic momenta (at constant positions) are:
\begin{equation}\label{eq:atomic_EOM_B1}
\frac{d \mathbf{p}_{i\alpha}}{dt} = - \dot\gamma \frac{m_{i\alpha}}{M_i} p_{yi} \hat x - \alpha_1 \frac{m_{i\alpha}}{M_i} \mathbf{p}_i.
\end{equation}
Here $\alpha_1$ is the part of the Gaussian thermostat multiplier $\zeta$ associated with conserving the kinetic energy during this part of the integration; it has the expression
\begin{equation}
\alpha_1 =\frac{ -\sum_i \dot\gamma p_{xi}p_{yi} /M_i} {\sum_i p_i^2/M_i}.
\end{equation}
Note that the thermostat term involves only molecular quantities, that is, it is
a so-called molecular thermostat which conserves the sum of the translational kinetic energies of the molecules $\sum_i p_i^2/2M_i$. Summing Eq.~\eqref{eq:atomic_EOM_B1} over $\alpha$ gives

\begin{equation}\label{eq:CM_EOM_B1}
\frac{d \mathbf{p}_{i}}{dt} = - \dot\gamma p_{yi} \hat x - \alpha_1 \mathbf{p}_i,
\end{equation}
which is identical to the $\mathbf{B_1}$ equation in Ref.~\onlinecite{pan} (where $i$ was an atomic index). Again the solution must be the same:

\begin{equation}
\mathbf{p}_i(t+\Delta t) = g(\Delta t) \left( \mathbf{p}_i(t)  - \hat x\Delta t \dot\gamma p_{yi}(t)\right).
\end{equation}
Here $g(\Delta t)=1/\sqrt{1-2c_1\Delta t + c_2\Delta t^2}$ and $c_1$ and $c_2$ are expressions involving sums of products of molecular momenta (see Ref.~\onlinecite{pan}). Next we consider the velocity relative to the center of mass. Dividing Eqs.~\eqref{eq:atomic_EOM_B1} and \ref{eq:CM_EOM_B1} by $m_{i\alpha}$ and $M_i$, respectively, and then subtracting the second from the first gives the equation for the relative velocity:

\begin{equation}
\frac{d}{dt} \left(\mathbf{v}_{i\alpha} - \mathbf{v}_i\right) = 0, 
\end{equation}
i.e. the relative velocity is not affected by the flow or the thermostat---this is of course the point of using molecular SLLOD equations and a molecular thermostat: they translate each molecules rigidly, without rotation or deformation. The update-expression for an atomic velocity is therefore that which gives  the same numerical difference as the update-expression for the center of mass velocity.

\begin{align}
\mathbf{v}_{i\alpha}(t+\Delta t) &= \mathbf{v_{i\alpha}}(t) -  \mathbf{v_{i}}(t) + \mathbf{v_{i}}(t+\Delta t) \\
&= \mathbf{v}_{i\alpha}(t) + (g(\Delta t)-1) \mathbf{v}_i (t) - g(\Delta t) \Delta t \dot\gamma v_{yi} \hat x.
\end{align}

For the $\mathbf{B2}$ equations of motion, the procedure is essentially the same: the velocity of the center of mass obeys the $\mathbf{B_2}$ equation in Ref.~\onlinecite{pan} for the atomic case, with solution

\begin{equation}\label{eq:sol_B2_CM}
\mathbf{p}_i(t+\Delta t) = \frac{1-h}{e-h/e} \left(\mathbf{p}_t(t) +
\mathbf{F}_i(t)\frac{1+h-e-h/e}{(1-h)\beta} \right) = \mathbf{p}_i(t) + M_i\Delta \mathbf{v}_i
\end{equation}
where expressions for $e(\Delta t)$, $\beta$ and $h$ are given in Ref.~\onlinecite{pan} (note that the sign of the second term in parentheses was incorrect in Ref.~\onlinecite{pan}, their Eq. (22)). The notation $\Delta \mathbf{v}_i$ for the change in center of mass velocity of molecule $i$ is introduced for use below. The relative velocity obeys:

\begin{equation}
\frac{d}{dt} \left(\mathbf{v}_{i\alpha} - \mathbf{v}_i\right) = \mathbf{F}_{i\alpha}/m_{i\alpha} - \mathbf{F}_{i}/M_{i} = \mathbf{a}_{i\alpha} -  \mathbf{a}_{i},
\end{equation}
where $\mathbf{a}_{i\alpha}$ and $\mathbf{a}_{i}$ are the (constant) accelerations.
To update the atomic velocities we then have

\begin{equation}
\mathbf{v}_{i\alpha}(t+\Delta t) =  \mathbf{v}_{i\alpha}(t) + \Delta v_i + \Delta t( \mathbf{a}_{i\alpha} -  \mathbf{a}_{i}),
\end{equation}
where $\Delta v_i $ is determined from Eq.~\eqref{eq:sol_B2_CM}.

A final technical point: while RUMD uses single precision floating point arithmetic in general, double precision is used for velocities in SLLOD algorithm. In this work, however, to keep the drift of the kinetic energy acceptably low for the small time step and strain rates used, double precision was used for the entire simulation code.

\end{document}